\documentclass[a4paper,12pt]{article}
\usepackage{amsfonts, amsmath, amssymb, amsthm}
\usepackage[toc,page]{appendix}
\usepackage[blocks]{authblk}
\usepackage{array}
\usepackage[english]{babel}
\usepackage{bm}
\usepackage{booktabs}
\usepackage{enumerate}
\usepackage{epsfig}
\usepackage{float}
\usepackage{fontenc}
\usepackage{graphicx}
\usepackage[hmargin=2cm,vmargin=3cm]{geometry}
\usepackage[colorlinks,allcolors = blue]{hyperref}
\usepackage{lscape}
\usepackage{mathrsfs}
\usepackage{multicol}
\usepackage{multirow,bigdelim}
\usepackage{natbib}
\usepackage{rotating}
\usepackage{subfigure}
\usepackage{verbatim}
\usepackage[svgnames]{xcolor}

\usepackage{mdframed}

\usepackage{color}

\def\red#1{{\color{red} #1}}%
\newtheorem{example}{Example}

\newcommand{\dif}{\ensuremath{\mathrm{d}}}

\newcommand{\T}{\ensuremath{\mathrm{\scriptscriptstyle T}}}
\usepackage{bm}

\newcommand{\alphab}{\ensuremath{\bm\alpha}}
\newcommand{\betab}{\ensuremath{\bm\beta}}
\newcommand{\gammab}{\ensuremath{\bm\gamma}}

\newcommand{\thetab}{\ensuremath{\bm\theta}}

\newcommand{\kappab}{\ensuremath{\bm\kappa}}

\newcommand{\psib}{\ensuremath{\bm\psi}}

\usepackage{amsmath}

\DeclareMathOperator{\E}{E}

\usepackage{mathrsfs}

\DeclareMathOperator{\Ga}{Gamma}

\DeclareMathOperator{\Lap}{Laplace}

 \let\oldthebibliography=\thebibliography
 \let\oldendthebibliography=\endthebibliography
 \renewenvironment{thebibliography}[1]{%
   \footnotesize
   \oldthebibliography{#1}%
   \setlength{\parskip}{0mm}%
   \setlength{\itemsep}{0mm}%
 }{\oldendthebibliography}

\def\EGPD{\text{EGPD}_G}

\usepackage{lscape}

\title{An Extreme Value Bayesian Lasso for \\ the Conditional {Left and Right Tails}}
\author[$\dag$]{M.~\textsc{de Carvalho}}
\affil[$\dag$]{School of Mathematics, University of Edinburgh, UK 
  (\href{mailto:miguel.decarvalho@ed.ac.uk}{miguel.decarvalho@ed.ac.uk})}
\author[$\ddag$]{S.~\textsc{Pereira}}
\affil[$\ddag$]{Faculdade de Ci\^encias and CEAUL, Universidade de Lisboa, Portugal
	(\href{mailto: sapereira@fc.ul.pt}{sapereira@fc.ul.pt})
}
\author[$*$]{P.~\textsc{Pereira}}
\affil[$*$]{ESTSetúbal/IPS and CEAUL, Portugal
(\href{mailto: paula.pereira@estsetubal.ips.pt}{paula.pereira@estsetubal.ips.pt})
}
\author[$\ddag$]{P.~\textsc{de Zea Bermudez}}

\date{}
\begin{document}
\maketitle 
\begin{abstract}\footnotesize 
  We introduce a novel regression model for the conditional
  {left and right tail} of a possibly heavy-tailed response. The proposed model can
be used to learn the effect of covariates on an extreme value setting 
via a Lasso-type specification based on a Lagrangian restriction. Our
model can be used to track if some covariates are significant for the
{lower values}, but not for the {(right)} tail---and vice-versa; in addition to this, the
proposed model bypasses the need for conditional threshold selection
in an extreme value theory framework. We assess the finite-sample
performance of the proposed methods through a simulation study that
reveals that our method recovers the true conditional distribution
over a variety of simulation scenarios, along with being accurate on
variable selection. Rainfall data are used to showcase how the
proposed method can learn to distinguish between key drivers of
moderate rainfall, against those of extreme rainfall.  \\

\noindent \textsc{key words:} Conditional tail; Extended Generalized Pareto
distribution; Heavy-tailed response; Lasso; $L_1$-Penalization;
Nonstationary extremes; Statistics of extremes; Variable selection.
\end{abstract}

\section{\large{\textsf{INTRODUCTION}}}\label{introduction} 
Learning about {the drivers} of risk is key {in} a variety of
fields, including climatology, environmental sciences,
finance, forestry, and hydrology.  Mainstream approaches for learning
about such drivers or predictors of risk include, for instance,
\cite{davison1990}, \cite{chavez-demoulin2005}, \cite{eastoe2009},
\cite{wang2009}, \cite{chavez-demoulin2016}, and \cite{huser2016}.

{The} main contribution of this paper rests on a Bayesian regression model
for the conditional {lower values (i.e.}{,}{~left tail)} and conditional {(right)} tail of a possibly heavy-tailed
response. Our model pioneers the development of regression methods in
an extreme value framework that allow for some covariates to be
significant for the {lower values} but not for the tail {(and vice-versa)}.  Some
further comments on the proposed model are in order. First, the
proposed model bypasses the need for (conditional) threshold
selection; such selection is particularly challenging {in} a regression
framework as it entails selecting a function of the covariate (say,
$u_{\mathbf{x}}$) rather than a single scalar. Second, our method models both
the conditional {lower values} and the conditional tail, {offering} a full portrait of the conditional distribution, while  still {being} able to extrapolate beyond observed data into the conditional
tail.  Finally, our method is directly tailored for variable selection
in an extreme value framework; in particular, it can be used to track
what covariates are significant for the {lower values} and tails.

In an extreme value framework, interest focuses on modelling the most
extreme observations{,} disregarding the central part of the
distribution. Usually, {efforts center on modelling} {extreme values using the} {generalized extreme value distribution or the generalized Pareto distribution}  \citep[][]{embrechts1997, coles2001,
  beirlant2004}. {Many} observations are disregarded using
the latter approaches, and the choice of the block size or threshold
are far from straightforward. Moreover, in many situations of applied
interest{,} it would be desirable to model both the {lower values} of the data
along with the extreme values in a regression framework. Our model
builds over \cite{papas2013} and \cite{naveau2016} who proposed an
{extended generalized Pareto distribution} (EGPD) {to jointly model}
low, moderate and extreme observations---without the need of threshold
selection; other interesting options for modeling both the bulk and
the tail of a distribution, include extreme value mixture models
\citep[e.g.{,}][]{frigessi2002, behrens2004, carreau2009, cabras2011,
  macdonald2011, nascimento2012} {as well as composition-based approaches  \citep{stein2020, stein2021}.}


The proposed model can be regarded as a Bayesian Lasso-type model for
the {lower values} and tail of a {possibly heavy-tailed response supported on the positive real line}.  The Bayesian Lasso was
introduced by \cite{park2008} as a Bayesian version of Tibshirani's
Lasso \citep{tibshirani1996}. Roughly speaking, the {frequentist}
Lasso is a regularization method, that shrinks some regression
coefficients, and sets others to {zero; as such it is   naturally tailored for variable selection}. {The frequentist
  Lasso solves a least squares type of optimization problem, where an}
  {$L_1$-penalization} {is incorporated via a constraint.} {The
  starting point for the Bayesian Lasso is the fact that the posterior
  mode generated from a Laplace prior coincides with the solution to
  the Lasso least squares optimization problem with} {$L_1$-penalization}
  {\citep[][Section~4.2.3]{reich2019}.  It should be noted however that
  the Bayesian Lasso does not set coefficients to zero.}

{Shrinkage in a Bayesian context has been traditionally achieved
  via priors with peaks at zero such as the Laplace.
  Other examples of continuous shrinkage priors include the Horseshoe
  \citep{carvalho2010}, Dirichlet--Laplace \citep{bhat2015}, and R2D2
  \citep{zhang2020}.} {There are also} {discrete shrinkage priors
  \citep{george1993} which are arguably more interpretable. Despite
  all these considerable advances in Bayesian variable selection, the
  Bayesian Lasso is simply one way to go, one that is associated}
  {$L_1$-penalization}{, but other priors could be used to meet a similar
  target, leading to other geometries for the corresponding
  penalization.} For a recent review of Bayesian regularization methods{,} see
\cite{polson2019}.

As we clarify below (Section~\ref{first}), our
model has also {links} with quantile regression. Indeed, by
modeling both the conditional {lower values} and the conditional tail, our model
bridges quantile regression \citep{koenker1978} with
extremal quantile regression \citep{chernozhukov2005}.  Finally, as a
byproduct, this paper contributes to the literature on Bayesian
inference for Pareto distributions 
\citep{arnold1989, de2003, castellanos2007, de2010, villa2017}.

The rest of this paper unfolds as follows. In Section~\ref{methods} we
introduce the proposed methods. {We assess} the
performance of the proposed methods {on simulated examples} {in Section~\ref{simulation}}, and report the main findings of
our numerical studies. A data illustration is included in
Section~\ref{application}{.} 

{\section{\large\textsf{EXTREME VALUE BAYESIAN LASSO}}\label{methods}}
To streamline the presentation, the proposed methods are introduced in
a step-by-step fashion, with the most flexible version of our model
being introduced in Section~\ref{extension}. 
{\subsection{\large\textsf{THE EXTENDED GENERALIZED PARETO FAMILY}}\label{egpd}}
Our starting point for modeling is the so-called extended generalized
Pareto distribution (EGPD), as proposed in \cite{papas2013} and
\cite{naveau2016}{; the EGPD is a distribution over the positive real line, whose} cumulative distribution function {is}
  $F(y) = {G\{H(y)\}}$, 
{where $G: [0, 1] \to [0, 1]$ is a carrier function, {which tilts} the generalized Pareto distribution (GPD) function}
\begin{equation*}
  H(y)=
  1-\left( 1+\frac{\xi y}{\sigma}\right)
  ^{-1 / \xi}, 
\end{equation*}
defined on $\{y \in (0, \infty): 1 + \xi y / \sigma > 0\}$.
Here, $\sigma>0$ is a scale parameter, $\xi \in \mathbb{R}$ is the shape
parameter; the case $\xi = 0$ should be understood
by taking the limit $\xi \to 0$. The
shape parameter $\xi$ is {termed} extreme value index and it
controls the rate of decay of the tail
. Following
\cite{naveau2016}, we assume that the carrier function $G$ obeys the following conditions:
\begin{itemize}
\item[A.] $\underset{v \to 0^+}{\lim}{\{1 - G(
	1-v)\}/v=a}$, with $a>0$. 
\item[B.] $\underset{v \to 0^+}{\lim} {G\left\lbrace v\,w(v)\right\rbrace /G(v)=b}$, with $b>0$ and $w(v) > 0$ {such that} $w(v)=1+o(v)$ as $v\rightarrow 0^+$. 
\item[C.] $\underset{v \to 0^+}{\lim}{G(v)/v^{\kappa}=c}$, with $c>0$.
\end{itemize}
Assumption~A ensures a Pareto-type tail, whereas Assumptions B--C
ensure that the {lower values are} driven by the carrier $G$. {In more detail,
  Assumption~A implies that
\begin{equation*}
  \lim_{y \to y_*} \frac{1 - F(y)}{1 - H(y)} = a, 
\end{equation*}
and thus it can be understood as a tail-equivalence
condition \citep[Section~3.3]{embrechts1997}, where $y^* = \inf\{y:
F(y) < 1\}$ is the so-called right endpoint, here assumed to be positive. 
Since tail-equivalence implies that both $F$ and $H$ are {in} the same
domain of attraction \citep[]{resnick1971}, it follows from Assumption~A
that $\xi$ can be literally interpreted as the extreme value index
of $F(y) = {G\{H(y)\}}$.} {We note further that Assumption~C implies that
small values follow a Weibull type GPD, and that the role and need of
Assumption B is in fact questionable \citep{tencaliec2019}; actually, the
latter paper makes no reference to it.}

For parsimony reasons{,} we focus on modeling $G$ using a parametric
class, so that $G(v) \equiv G_{\kappab}(v)$, with
$\kappab \in \mathbb{R}^q$. The canonical example of a parametric
carrier is $G_\kappa(v) = v^{\kappa}$, with $\kappa > 0$
{controlling} the shape of the lower tail, with a larger value of
  $\kappa$ leading to less mass close to zero; we refer to the EGPD
  distribution with the latter carrier as the canonical EGPD.

Below, we use the notation $Y \sim \EGPD(\kappab, \sigma, \xi)$ {to}
denote that $Y$ follows an EGPD with parameters $(\kappab, \sigma, \xi)$
{and} carrier $G${, and $\mathscr{G}$ to}  represent the
space of all carrier functions $G: (0, \infty) \to [0, 1]$ obeying
Assumptions~{A--C}.

\subsection{\large\textsf{A FIRST CONDITIONAL MODEL FOR THE {LOWER VALUES} AND TAIL OF A POSSIBLY HEAVY-TAILED RESPONSE}}\label{first}
The first version of our model {specifies} the conditional distribution function
\begin{equation}\label{model}
  F(y \mid \mathbf{x}) = G_{\kappab(\mathbf{x})}(H(y)), 
\end{equation}
where $\mathbf{x} = (x_1, \dots, x_p) \in \mathbb{R}^p$ is a vector of covariates {and} $\kappab(\mathbf{x})$ is a {vector-valued} function with components given by inverse link functions
\begin{equation}\label{links0}
  \kappab(\mathbf{x}) ={ [\kappa_1(\mathbf{x}^{\T}\betab_1), \dots, \kappa_q(\mathbf{x}^{\T}\betab_q)],}
\end{equation}
and $G_{\kappab(\mathbf{x})} \in \mathscr{G}$, for every $\mathbf{x}
\in \mathcal{X} \subseteq \mathbb{R}^p$; here and below, $\betab_j =
(\beta_{1, j}, \dots, \beta_{p, j}) \in \mathbb{R}^p$, for $j = 1,
\dots, q$.
For reasons that will become evident {later}, {we do not allow for now} $(\sigma, \xi)$ to depend on the {covariates} $\mathbf{x}$, but we will extend the specification in \eqref{model} {in Section}~\ref{extension}. {The next examples} {illustrate} {that different types of carrier functions exist, and {thus that} there exist different forms of tilting the GPD distribution function via a carrier function; the examples also highlight that, given the freedom to choose a $G$ obeying Assumptions~A--C, the model is quite flexible and that thus it would be challenging to numerically illustrate all its instances.}
\begin{example}[Power carrier, exponential link]\label{powerc}\normalfont 
  The canonical embodiment of the first version of our model in \eqref{model} is
  obtained by specifying
  \begin{equation}\label{Gx}
    G_{\kappa(\mathbf{x})}(v) = v^{\kappa(\mathbf{x})}, \quad \kappa(\mathbf{x}) = \exp(\mathbf{x}^{\T} \betab).
  \end{equation}
\end{example}
\begin{example}[Beta carrier, exponential link]\label{betac}\normalfont 
  Another variant of \eqref{model} is obtained by specifying
  \begin{equation}\label{Gx}
    G_{\kappa(\mathbf{x})}(v) = 1 - Q_{\kappa(\mathbf{x})}(1 - v^{\kappa(\mathbf{x})}), \quad \kappa(\mathbf{x}) = \exp(\mathbf{x}^{\T} \betab),
  \end{equation}
  where 
  \begin{equation*}
    Q_{\kappa(\mathbf{x})}(v) = \frac{1 + \kappa(\mathbf{x})}{\kappa(\mathbf{x})} v^{1 / \kappa(\mathbf{x})} \bigg(1 - \frac{v}{1 + \kappa(\mathbf{x})} \bigg)
  \end{equation*}
is the distribution function of a Beta distribution with parameters $(1 / \kappa(\mathbf{x}), 2)$. 
\end{example}
\begin{example}[Power mixture carrier, exponential links]\label{powerm}\normalfont 
Still another variant of {\eqref{model}} is obtained by specifying 
\begin{equation*}
  G_{\kappab(\mathbf{x})}(v) = \pi v^{\kappa_1(\mathbf{x})} + 
  (1 - \pi)  v^{\kappa_2(\mathbf{x})},
\end{equation*}
with $0 < \pi < 1$ and 
\begin{equation*}
  \kappa_1(\mathbf{x}) = \exp(\mathbf{x}^{\T} \betab_1), \quad 
  \kappa_2(\mathbf{x}) = \exp(\mathbf{x}^{\T} \betab_2).
\end{equation*}
The conditions on the intercepts $\beta_{1,1} > \beta_{1,2}$ and $\beta_{j,1} =
\beta_{j,2}$, for $j = 2, \dots, q$, leads to $\kappa_1(\mathbf{x}) >
\kappa_2(\mathbf{x})$ and is used for {identifiability} purposes.
\end{example}

\noindent A consequence of \eqref{model} is that
\begin{equation} \label{qr2}
F^{-1}(p \mid \mathbf{x}) = 
\begin{cases}
\frac{\sigma}{\xi} [\{1 - G^{-1}_{\kappab(\mathbf{x})}(p)\}^{-\xi} - 1], & \xi \neq 0,\\
- \frac{\sigma}{\xi} \log\{1 - G^{-1}_{\kappab(\mathbf{x})}(p)\}, & \xi = 0,
\end{cases}
\end{equation}
where $F^{-1}(p \mid \mathbf{x}) = \inf\{y: F(y \mid \mathbf{x}) \geq p\}$, for $0 < p < 1$. Equation~\eqref{qr2} warrants some remarks on links with quantile regression. The first version of our model in \eqref{model} can be regarded as a model that bridges quantile regression \citep{koenker1978} with extremal quantile regression \citep{chernozhukov2005}, in the sense that it offers a way to model both moderate and high quantiles. Quantile regression \citep{koenker2005} allows for each $\tau$th conditional quantile to have its own slope $\betab_\tau$, according to the following linear specification
\begin{equation}\label{qr}
  F^{-1}(\tau \mid \mathbf{x}) = \mathbf{x}^{\T}\betab_\tau,  
  \quad 0 < \tau < 1.
\end{equation}
In the same way that high empirical quantiles fail to 
extrapolate into the tail of a distribution, the standard version of
quantile regression in \eqref{qr} is unable to extrapolate into the
tail of the conditional response.

{We describe next} Bayesian Lasso modeling and {tackle} inference for the first version of our model.
For parsimony, below we will focus on the version of the model that sets $q = 1$ in \eqref{links0}, so that $\kappa(\mathbf{x}) = {\kappa_1}(\mathbf{x}^{\T} \betab)$. {We underscore that what is studied in the next section applies with some minor modifications to the full model presented above; the focus on $q = 1$ (to which Examples 1~2 apply) in the next section eases however notation, and it obviates the need for discussing again the {identifiability} issues that we have alluded to already in Example~3.}

\subsection{\large\textsf{REGULARIZATION AND BAYESIAN INFERENCE}}\label{inference}
Let $\{(\mathbf{x}_i, y_i)\}_{i = 1}^n$ be a random sample from
$F(\mathbf{x}, y)$. We propose a Bayesian Lasso-type specification
for the model in \eqref{model} so {as} to regularize the
log likelihood. Specifically, let $G_{\kappa(\mathbf{x})} \in \mathscr{G}$, with $\kappa(\mathbf{x})$ being a function of $\mathbf{x}$, so that the likelihood becomes 
\begin{equation*} 
  L(\betab, \sigma, \xi) = \prod_{i = 1}^n h(y_i) g_{\kappab(\mathbf{x}_i)}{\{H(y_i)\}},
\end{equation*}
where $\betab = (\betab_1, \dots, \betab_q)$, $h(y) = {\sigma^{-1}} (1 + \xi y / \sigma)_{{+}}^{-1/\xi - 1}$ is the density of the generalized Pareto distribution, $g_{\kappab(\mathbf{x})} = \dif G_{\kappab(\mathbf{x})} / \dif y$, and {$(a)_{+} = \max(0, a)$ is the positive part function.} 
In a Bayesian context{,} {the constrained optimization problem underlying the Lasso can be equivalently rewritten as the} posterior mode {of regression parameters, provided that one assumes that those} {are (a priori)} independent {with} identical Laplace priors (i.e.{,}~double exponential), that
is, {$\pi(\betab) \propto \prod_{j=1}^{q} \text{e}^{-\lambda / 2 \, |\beta_j|}$}; see \cite{tibshirani1996} and \citet[][Section~4.2.3]{reich2019}. Following \cite{park2008}, we assume a Gamma prior on $\lambda^2$. The hierarchical representation of the first version of our Bayesian Lasso conditional EGPD for an heavy tailed response is thus: 
\begin{mdframed}
  \textbf{Extreme Value Bayesian Lasso for the Conditional {Lower Values} and Tail}\\
  (1st version)
  \begin{enumerate}   
  \item \textbf{Likelihood} \vspace{-0.4cm}
    \begin{equation*}
      y_i \mid \mathbf{x}_i,\betab,\sigma, \xi, \lambda^2 \sim {\EGPD[\kappa(\mathbf{x}_i), \sigma, \xi]}, \quad i = 1, \dots, n,
    \end{equation*}
    \begin{equation}\label{links}
      \kappa(\mathbf{x}) = k(\mathbf{x}^{\T}\betab).
    \end{equation}
    \vspace{-1cm}  
  \item \textbf{Priors} \vspace{-0.2cm}
    \begin{equation*}
      \begin{cases}
      \begin{split}
        {\betab_j} &\mid \lambda \overset{\text{iid}}{\sim} \Lap(\lambda), \\
        \lambda^2 &\sim \Ga(a_{\lambda},b_{\lambda}), \\
        \sigma &\sim \Ga(a_\sigma,b_\sigma), \\
        \xi &\sim \, \pi_\xi.
      \end{split}
    \end{cases}
  \end{equation*}   
\end{enumerate}
\end{mdframed}
{When the goal is to set a prior on the space of heavy-tailed distributions, then it may be sensible} to set 
$\pi_\xi = \text{Gamma}(a_\xi, b_\xi)$, whereas $\pi_\xi = \text{N}(\mu_\xi, \sigma_\xi)$ may be sensible if all domains attraction are equally likely a priori. Since the posterior has no closed{-}form expression{,} we resort to {Markov Chain
Monte Carlo (MCMC)} methods to sample {from the posterior. }

 
{A} shortcoming of the first version of the {model} is that it only allows for the effect
of covariates on the {lower values}, but not on the tail; this is a consequence
of the fact that only the part of the model that drives the {lower values},
$\kappa(\textbf{x})$, is indexed by a covariate. {This issue is
 addressed in the next section.} 

\subsection{\large\textsf{EXTENSIONS FOR COVARIATE-ADJUSTED TAIL}}\label{extension}
In practice some covariates can be significant for the {lower values} but not
for the tail---or the other way around. Thus, we extend the specification from
Sections~\ref{first}--\ref{inference} by also allowing parameters
underlying the tail {to depend on covariates}. Specifically, we
consider the following specification:
\begin{equation}\label{egpdx}	
F(y \mid \mathbf{x})=G_{\kappa(\mathbf{x})}{\{H(y \mid \mathbf{x})\}},
\end{equation}
where $H(y \mid \mathbf{x})$ is a reparametrized conditional
generalized Pareto distribution, with parameters $\nu(\mathbf{x}) = \sigma(\mathbf{x}) \{1 + \xi(\mathbf{x})\}$ and $\xi(\mathbf{x})$, that is  
\begin{equation*}
  H(y \mid \mathbf{x})= 1-\left[ 1+\frac{\xi(\mathbf{x})\{1 + \xi(\mathbf{x})\} y}{\nu(\mathbf{x})}\right]^{-1 / \xi}.
\end{equation*}
Here, $\kappa(\mathbf{x})$ is a function as in \eqref{links}
whereas $\nu(\mathbf{x}) = {\phi}(\mathbf{x}^{\T}\alphab)$ and $\xi(\mathbf{x}) =
\mu(\mathbf{x}^{\T}\gammab)$, with {$\phi$} and $\mu$ being inverse-link functions.
The canonical embodiment of the full version of our model is obtained by
specifying $G_{\kappa(\mathbf{x})}(v) = v^{\kappa(\mathbf{x})}$ along
with
\begin{equation*}\label{Gx2}
  \kappa(\mathbf{x}) = \exp(\mathbf{x}^{\T} \betab), \quad 
  \nu(\mathbf{x}) = \exp(\mathbf{x}^{\T} \alphab), \quad 
  \xi(\mathbf{x}) = {\exp(\mathbf{x}^{\T} \gammab)},
\end{equation*}

\noindent {assuming $\xi(\mathbf{x}) > 0$.} The schematic representation below summarizes our model:
\begin{mdframed}
  \textbf{Extreme Value Bayesian Lasso for the Conditional {Lower Values} and Tail}\\
  (2nd version)
  \begin{enumerate}   
  \item \textbf{Likelihood} \vspace{-0.4cm}
    \begin{equation*}
      y_i \mid \mathbf{x}_i,\alphab, \betab, \gammab, \lambda_\alpha \sim \EGPD{[}\kappa(\mathbf{x}_i), \sigma(\mathbf{x}_i), \xi(\mathbf{x}_i){]}, \quad i = 1, \dots, n,
    \end{equation*}
    \begin{equation}\label{links2}
    {
        \kappa(\mathbf{x}) = k(\mathbf{x}^{\T} \betab), \quad 
        \nu(\mathbf{x}) = \ell(\mathbf{x}^{\T} \alphab), \quad
        \xi(\mathbf{x}) = \mu(\mathbf{x}^{\T} \gammab).\\
       }
    \end{equation}
    \vspace{-1cm}  
  \item \textbf{Priors}
    \begin{equation*}
     {
      \begin{cases}
      \begin{split}        
        {\beta_i} &\mid {\lambda_\beta} \overset{\text{iid}}{\sim} \Lap({\lambda_\beta}), \quad         {\lambda_\beta \sim \Ga(a_\beta, b_\beta)},      \\
        {\alpha_i} &\mid {\lambda_\alpha} \overset{\text{iid}}{\sim} \Lap({\lambda_\alpha}), \quad        {\lambda_\alpha \sim \Ga(a_\alpha, b_\alpha)},      \\
        {\gamma_i} &\mid {\lambda_\gamma} \overset{\text{iid}}{\sim} \Lap({\lambda_\gamma}), \quad         {\lambda_\gamma \sim \Ga(a_\gamma, b_\gamma)}.\\      
      \end{split}
      \end{cases}
    }
    \end{equation*} 
  \end{enumerate}
\end{mdframed}
\noindent The specification in \eqref{egpdx} warrants some remarks: 
\begin{enumerate}
\item \textit{The need for a reparametrization}: While it would seem
  natural to simply index $(\sigma, \xi)$ with a covariate and to
  proceed as in Sections~\ref{first}--\ref{inference}, similarly to
  \cite{chavez-demoulin2005} and \cite{cabras2011}, who work in a GPD
  setting we found that approach to be computationally unstable; in
  particular, in our case the latter parametrization leads to poor
  mixing and to small effective sample sizes.
\item \textit{Any {potential} reparametrization should keep parameters
    for the {lower values} and tails separated}: Since we aim to learn which covariates
  are important for the {lower values} and tail, any reparametrization to be made
  should not mix the parameters for the conditional {lower values} and tail, i.e., we look for
  reparametrizations of the type
  \begin{equation*}
    (\kappa, \sigma, \xi) \mapsto {\{}\kappa, \nu(\sigma, \xi), \xi{\}}
    \quad \text{or} \quad
    (\kappa, \sigma, \xi) \mapsto {\{}\kappa, \sigma, \zeta(\sigma, \xi){\}}.
  \end{equation*}
\item \textit{Fisher information}: Parameter orthogonality
  \citep[][Section~9.2]{young2005} requires the computation of the
  Fisher information matrix.  {The} canonical EGPD
  is a particular case of the so-called {generalized} Feller--Pareto distribution \citep{kleiber2003}, with $(a, b, p, q, r) = (1, \sigma / \xi, \kappa, 1, 1 / \xi)$, as 
  \begin{equation*}
    h_{\text{GFP}}(y) = \frac{a r y^{a - 1}}{b^a B(p, q)} \left\{1 + \left(\frac{y}{b}\right)^a\right\}^{-r q - 1}
    \left[1 - \left\{1 + \left(\frac{y}{b}\right)^a\right\}^{-r} \right]^{p - 1}, \quad y > 0,
  \end{equation*}
  where $B(p, q) = \int_0^1 t^{p - 1} (1 - t)^{q - 1} \, \dif t$, 
  the entries of the Fisher information matrix follow from
  \citet[][Section~3]{mahmoud2015}; yet the matrix is rather intricate
  and thus we decided to look for a more sensible way to proceed than 
  attempting to achieve parameter orthogonality.  
\end{enumerate}
The reparametrization $(\sigma, \xi) \mapsto (\sigma(1 + \xi), \xi)$
is orthogonal for the case of the GPD {for $\xi > - 1 / 2$} \citep{chavez-demoulin2005}, but it is only approximately
orthogonal for the EGPD in a neighborhood of $\kappa = 1$. Heuristically, this can
be seen by contrasting the likelihood of the canonical EGPD ($l$) with
that of the GPD ($l^*$); to streamline the argument, we concentrate on the single
observation case $(y)$ but the derivations below {holds} more generally. The
starting point is that
\begin{equation}\label{proxl}
  l(\psib) = l(\kappa, \thetab) = l^*(\thetab) + \log \kappa + (\kappa - 1) \log\{H(y)\}, 
\end{equation}
{where} $\psib= (\sigma, \xi, \kappa)$ and $\thetab = (\sigma, \xi)$, 
and thus it follows from \eqref{proxl} that ${l(\psib)} \approx l^*(\thetab)$, in a neighborhood of $\kappa = 1$; a similar relation holds for the corresponding Fisher informations,
\begin{equation*}
  I_{\psib} = -\E_{\psib} \left(\frac{\partial^2 l}{\partial \psib \partial \psib^{\T}} \right)
  = \left(
    \begin{matrix}
      I_{\thetab} & I_{\thetab, \kappa} \\
      I_{\kappa, \thetab} & I_{\kappa}      
    \end{matrix}
  \right), \quad
  I_{\thetab}^* = -\E_{\thetab} \left(\frac{\partial^2 l^*}{\partial \thetab \partial \thetab^{\T}} \right),
\end{equation*}
where $\E_{\thetab}(\cdot) = \int \cdot ~h_{\thetab}(y) \, \dif y$ and  $\E_{\psib}(\cdot) = \int \cdot ~h_{\thetab}(y) g{\{}H_{\thetab}(y){\}}\, \dif y$. {Indeed,} matrix calculus \citep[e.g.,][Chapter~9]{schott2016} can be used to show that that
\begin{equation*}
  I_{\thetab} = -\E_{\psib}\left(\frac{\partial^2 l^*}{\partial \thetab \partial \thetab^{\T}}\right) - (\kappa - 1) \E_{\psib}(M_{\thetab}),
\end{equation*}
where 
\begin{equation*}
  M_{\thetab} =  \left(H \frac{\partial^2 H}{\partial \thetab \partial \thetab^{\T}} - \frac{\partial H}{\partial \thetab} \frac{\partial H}{\partial \thetab^{\T}}\right) {H^{-2}}, \quad H \equiv H(y),
\end{equation*}
thus suggesting that $I_{\thetab} \approx I_{\thetab}^*$ in a neighborhood
of $\kappa = 1$. Despite the fact the reparametrization is only
quasi-orthogonal for $(\sigma, \xi)$ in an EGPD framework,
{numerical studies} reported in Section~\ref{simulation} indicate very
good {performances in terms of accuracy of fitted 
  regression estimates and QQ-plots of randomized quantile residuals \citep{dunn1996}}.


\section{\large\textsf{SIMULATION STUDY}}\label{simulation}
\subsection{\large\textsf{SIMULATION SCENARIOS AND ONE-SHOT EXPERIMENTS}}\label{oneshot}
In this section{,} we describe the simulation scenarios and present one-shot experiments so {as} to illustrate the proposed method; a Monte Carlo study {is} presented in Section~\ref{mc}. For now, we focus on describing the setting over which we simulate the data and on discussing the numerical experiments. {Code for implementing our methods is available from the Supplementary Material; we used \texttt{jags} \citep{plummer2019} as we have more experience with it, but \texttt{stan} \citep{gelman2015} would look like another natural option that could potentially yield faster mixing and more effective sample sizes at little extra programming cost.} {Throughout we use uninformative Gamma priors, Gamma(0.1, 0.1), for the hyperparameters $\lambda_k$, $\lambda_\nu$, and $\lambda_{\xi}$.}

Data are simulated according to \eqref{egpdx}, with power carrier  $G_{\kappa(\mathbf{x})}(v) = v^{\kappa(\mathbf{x})}$ and with link functions
\begin{equation*}
  \kappa_{\mathbf{x}} = \exp(\mathbf{x}^{\T} \betab), \quad
  \nu_{\mathbf{x}} = \exp(\mathbf{x}^{\T} \alphab), \quad 
  {
  \xi_{\mathbf{x}} = \exp(\mathbf{x}^{\T} \gammab).
}
\end{equation*}

\begin{figure}[H]\centering
 \textbf{Scenario~1} \hspace{5.3cm} \textbf{Scenario~2} \hspace{1.5cm} \\
 \footnotesize (light effects for {lower values}, light effects for tail) \hspace{0.5cm} (light effects for {lower values}, large effects for tail)\\
  \includegraphics[width=0.465\textwidth]{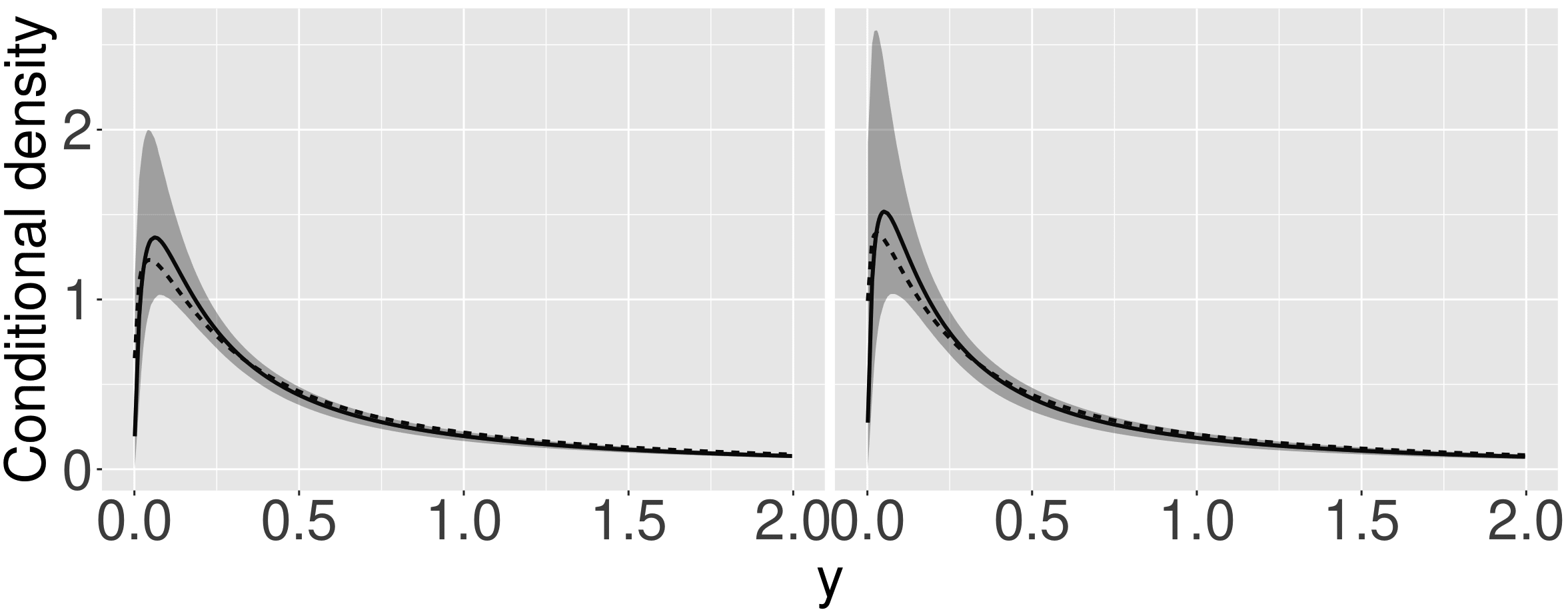}
  \includegraphics[width=0.465\textwidth]{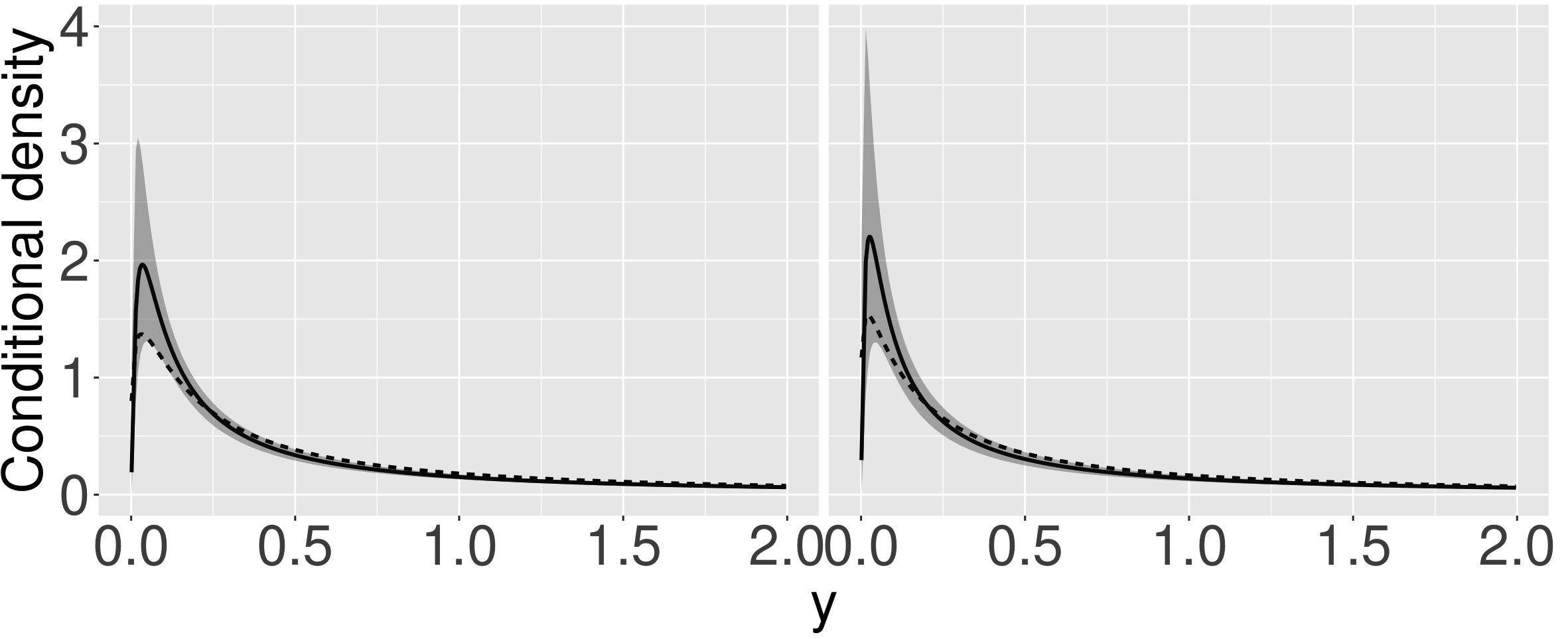}\\
  \textbf{Scenario~3} \hspace{5.3cm} \textbf{Scenario~4} \hspace{1.5cm} \\ \centering 
  (large effects for {lower values}, light effects for tail) \hspace{0.5cm} (large effects for {lower values}, large effects for tail)\\
  \includegraphics[width=0.465\textwidth]{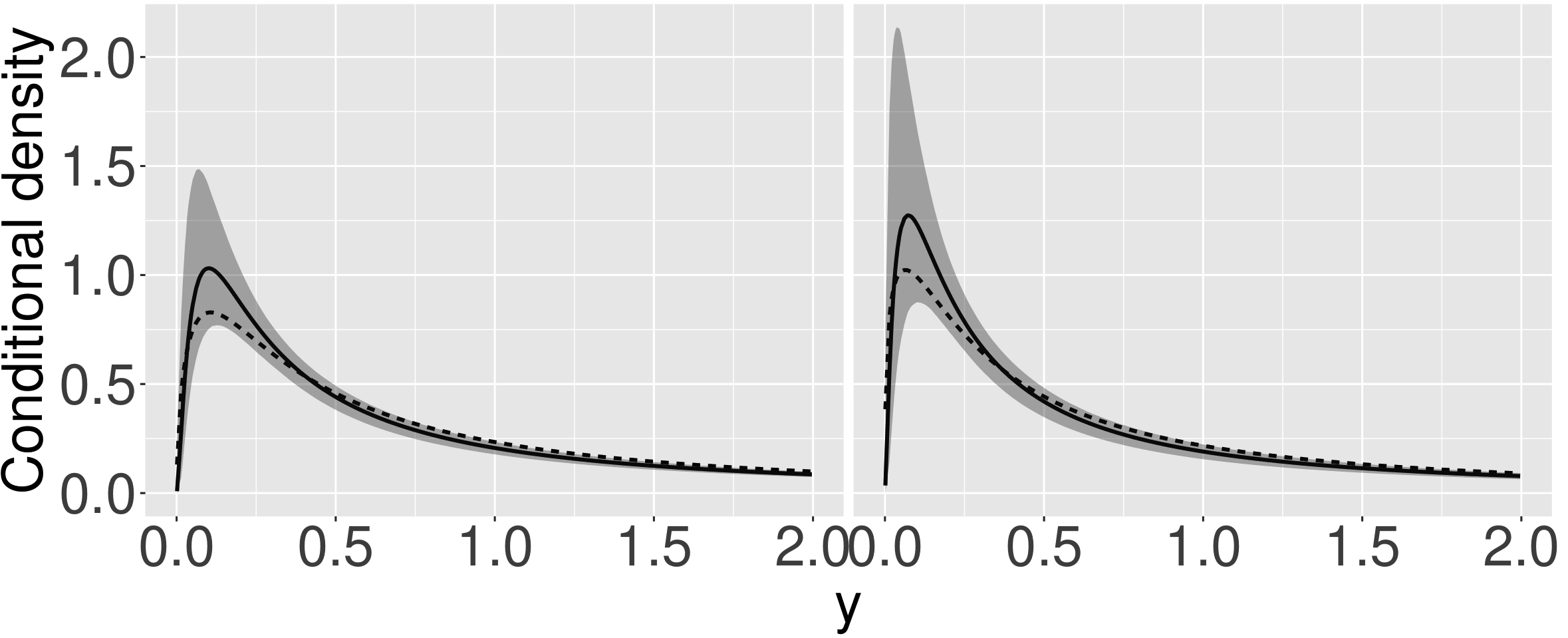}
  \includegraphics[width=0.465\textwidth]{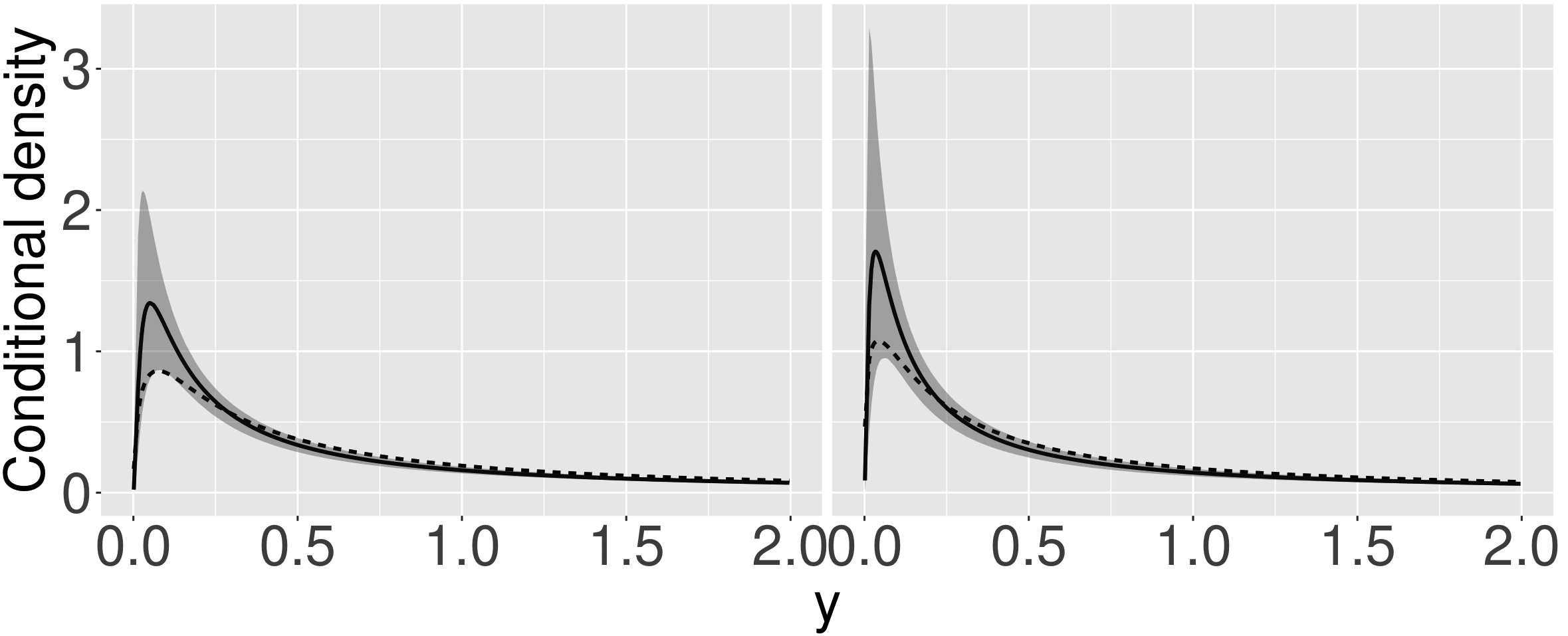}
  
  \caption{\label{scenariosim} \footnotesize Cross sections of
    {posterior mean conditional density} (solid) along with {pointwise} credible bands against true (dashed) for a one-shot experiment with $n = 500$; the cross sections result from
    conditioning on $\mathbf{x} = (0.25, \dots, 0.25)$ (left) and $\mathbf{x} = (0.50,
    \dots, 0.50)$ (right).
}
\end{figure}

\noindent {We consider four scenarios, described below, based on the following vectors for the regression coefficients:}

\begin{itemize}
\item \textbf{Scenario~1}---Light effects for {lower values}, light effects for tail: $\betab=\mathbf{a}; \alphab=\mathbf{b}; \gammab=\mathbf{c}$.
\item \textbf{Scenario~2}---Light effects for {lower values}, large effects for tail: $\betab=\mathbf{a}; \alpha=2\mathbf{b}; \gamma=2\mathbf{c}$
\item \textbf{Scenario~3}---Large effects for {lower values}, light effects for tail: $\betab=2\mathbf{a}; \alphab=b; \gammab=\mathbf{c}$
\item \textbf{Scenario~4}---Large effects for {lower values}, large effects for tail: $\betab=2\mathbf{a}; \alpha=2\mathbf{b}; \gamma=2\mathbf{c}$
\end{itemize}
where {$\mathbf{a}$, $\mathbf{b}$, $\mathbf{c} \in \mathbb{R}^{10}$ have components }
{
\begin{equation*}
            a_1 = a_3 = 0.3; a_6 = a_{10} = -0.3; \quad 
            b_2 = -0.3; b_5 = b_8 = 0.3; \quad 
            c_1 =  c_4 = c_9 = 0.3; c_{10} = -0.3;
\end{equation*}}
{and zero otherwise}.


For each of the scenarios above, we {simulated} $n = 500$ observations, $\{(\mathbf{x}_i, y_i)\}_{i = 1}^n$, from a conditional EGPD with a canonical carrier function; this is about the same number of observations {available} in the data application {of} Section~\ref{application}. The covariates are simulated from independent standard uniform distributions. Figure~\ref{scenariosim} shows cross sections of the true and conditional densities for Scenarios~1--4  estimated using our {methods.}

As {can be} seen from these figures{,} the method recovers satisfactorily well the true conditional density---especially keeping in mind {the sample size}.

\begin{figure}
  \begin{minipage}{0.48\linewidth}\hspace{1cm}
    \textbf{Scenario~1}\\ \footnotesize 
    (light effects for {lower values}, light effects for tail) \\
		\centering 
			\includegraphics[scale = .07]{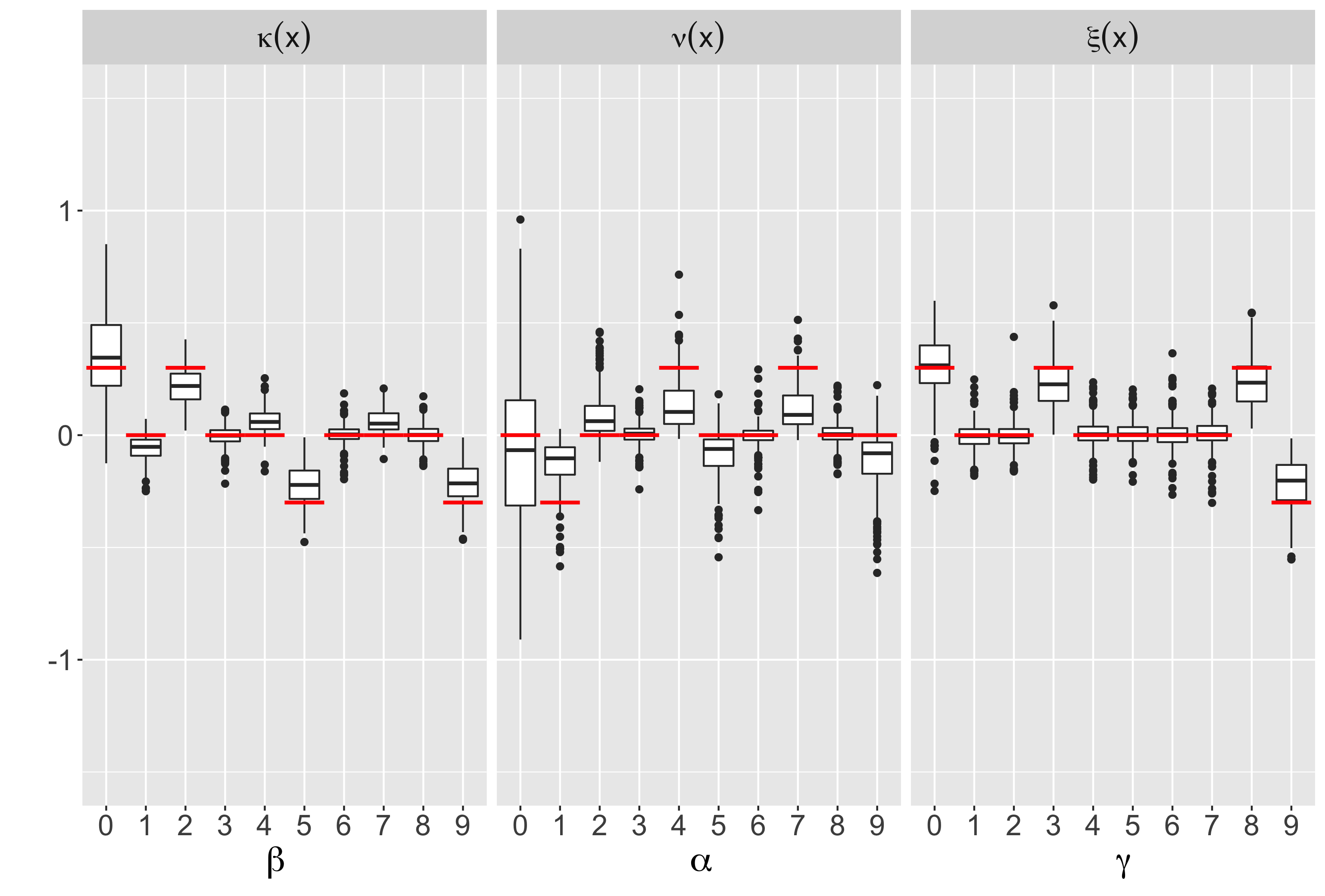}
              \end{minipage}
              \begin{minipage}{0.48\linewidth}\hspace{1cm}
                \textbf{Scenario~2}\\ \footnotesize 
       (light effects for {lower values}, large effects for tail)\\
		\centering
		\includegraphics[scale = .07]{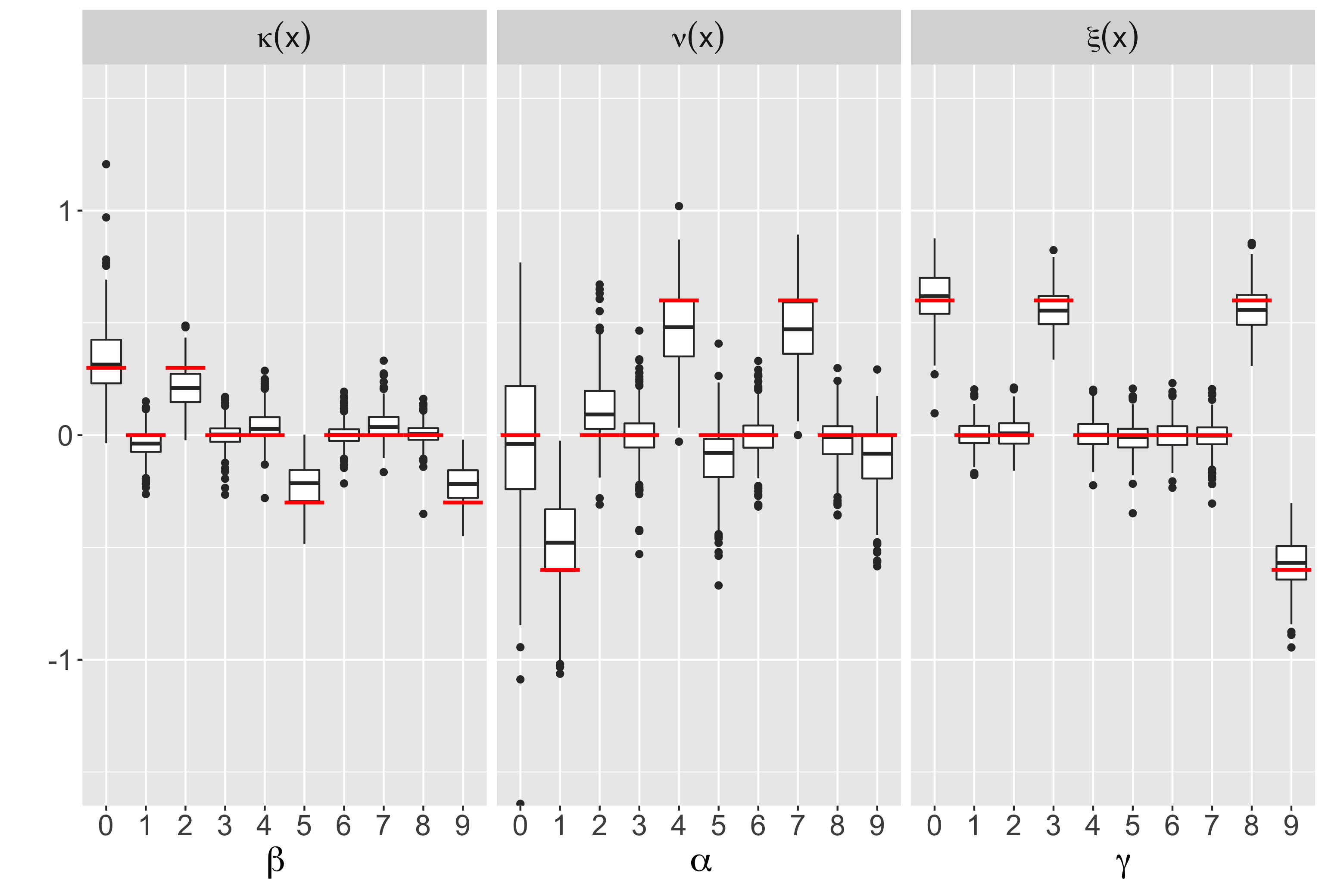}
              \end{minipage}\\
              \begin{minipage}{0.48\linewidth}\hspace{1cm}
                \textbf{Scenario~3}\\ \footnotesize 
                 (large effects for {lower values}, light effects for tail)\\
		\centering
			\includegraphics[scale = .07]{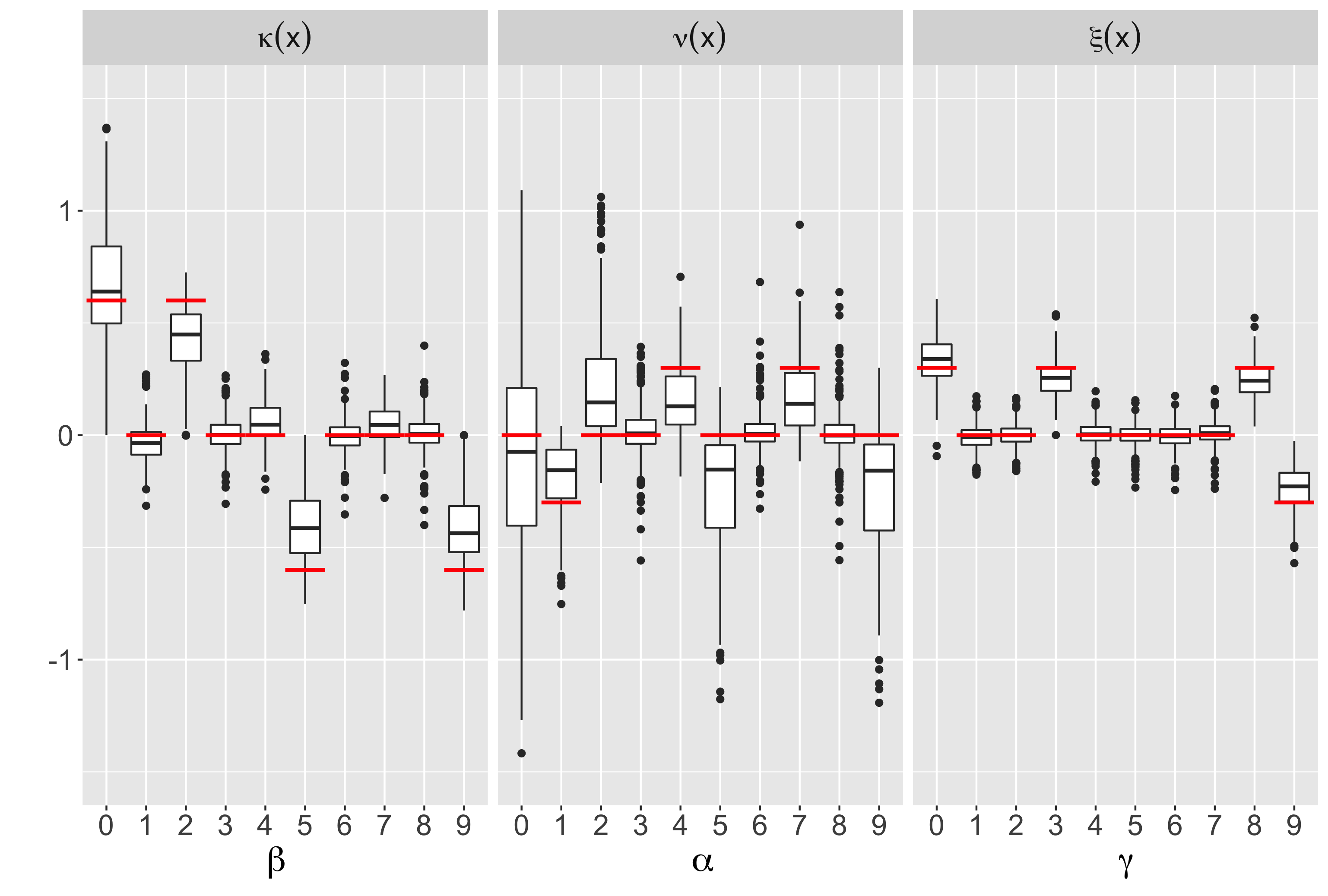}
              \end{minipage}
              \begin{minipage}{0.48\linewidth}\hspace{1cm}
                \textbf{Scenario~4}\\ \footnotesize 
                 (large effects for {lower values}, large effects for tail)\\
		\centering
		\includegraphics[scale = .07]{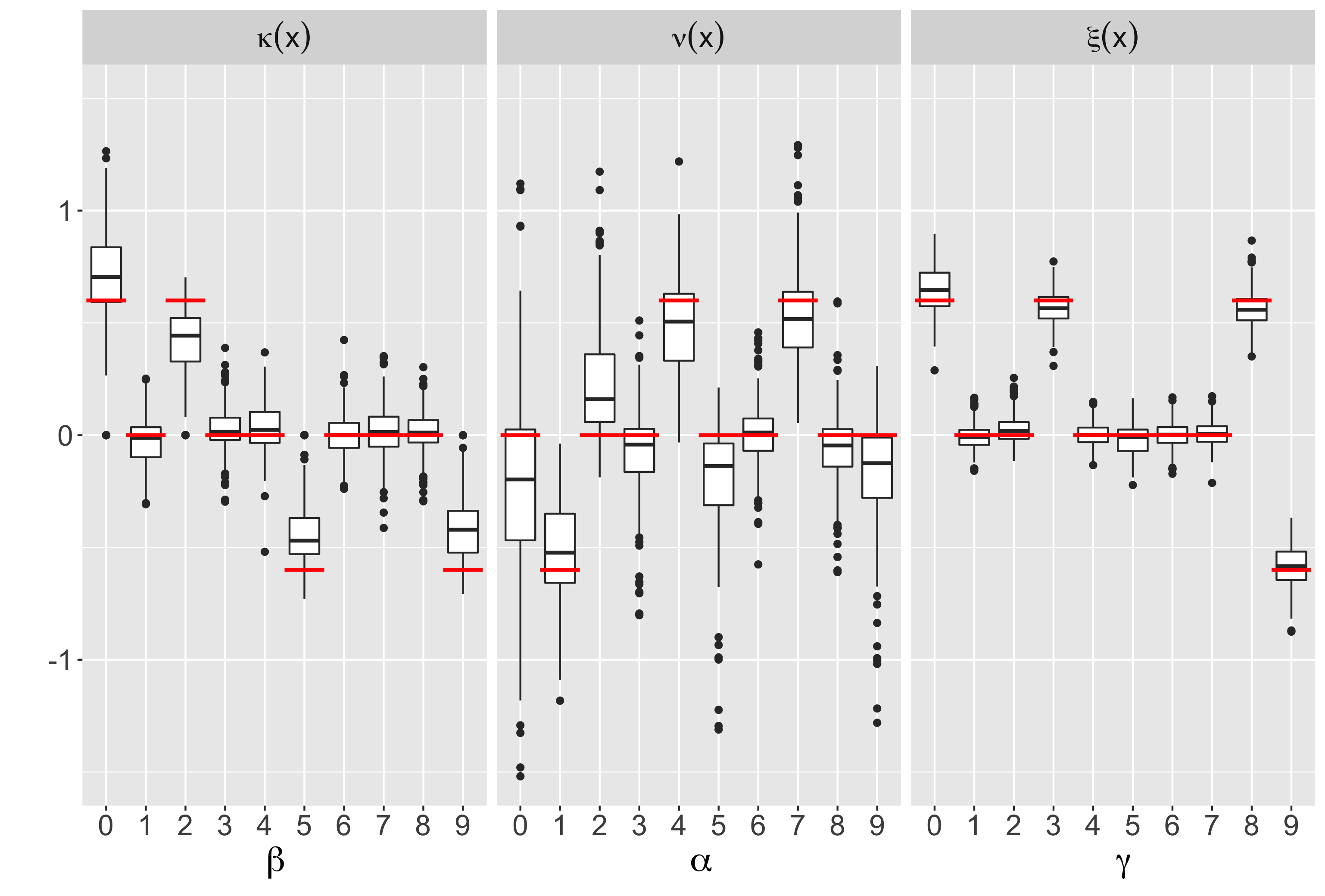}
	\end{minipage}
	\caption{\label{boxplots} \footnotesize Side-by-side boxplots with regression coefficient estimates for Monte Carlo simulation study ($n = 250$) plotted against the true values (\red{---}). The values 1--9 represent the coefficient indices.}
\end{figure}

\subsection{\large\textsf{MONTE CARLO SIMULATION STUDY}}\label{mc}
To assess the finite-sample performance of the proposed methods{,} we now present the results of a Monte Carlo simulation study, where we repeat 250 times the one-shot experiments from Section~\ref{oneshot}. We consider the {different} sample sizes {of} $n = 100$, $n = 250$, and $n = 500$.

{Figure~\ref{boxplots} presents side-by-side boxplots of the coefficient estimates for each scenario for the case $n = 250${, and we can see that} the estimates tend to be close to the true values thus suggesting that the proposed methods are able to learn what covariates are significant for the {lower values}, but not for the tail (and vice-versa); further {simulation} experiments (results not shown) indicate that performance may deteriorate if the number of common effects between {left and right tails} is large. The coefficient estimates for the cases $n = 100$ and $n = 500$ are reported in the Supplementary Material{;} as expected{,} the larger the sample size{,} the more accurate the estimates.} {The frequency of variable selection table presented in the Supplementary Material suggests a satisfactory performance of the method in terms of variable selection.}

\begin{figure}
  \centering
  \textbf{Scenario~1}\\
   (light effects for {lower values}, light effects for tail)\\
\includegraphics[width=0.45\textwidth]{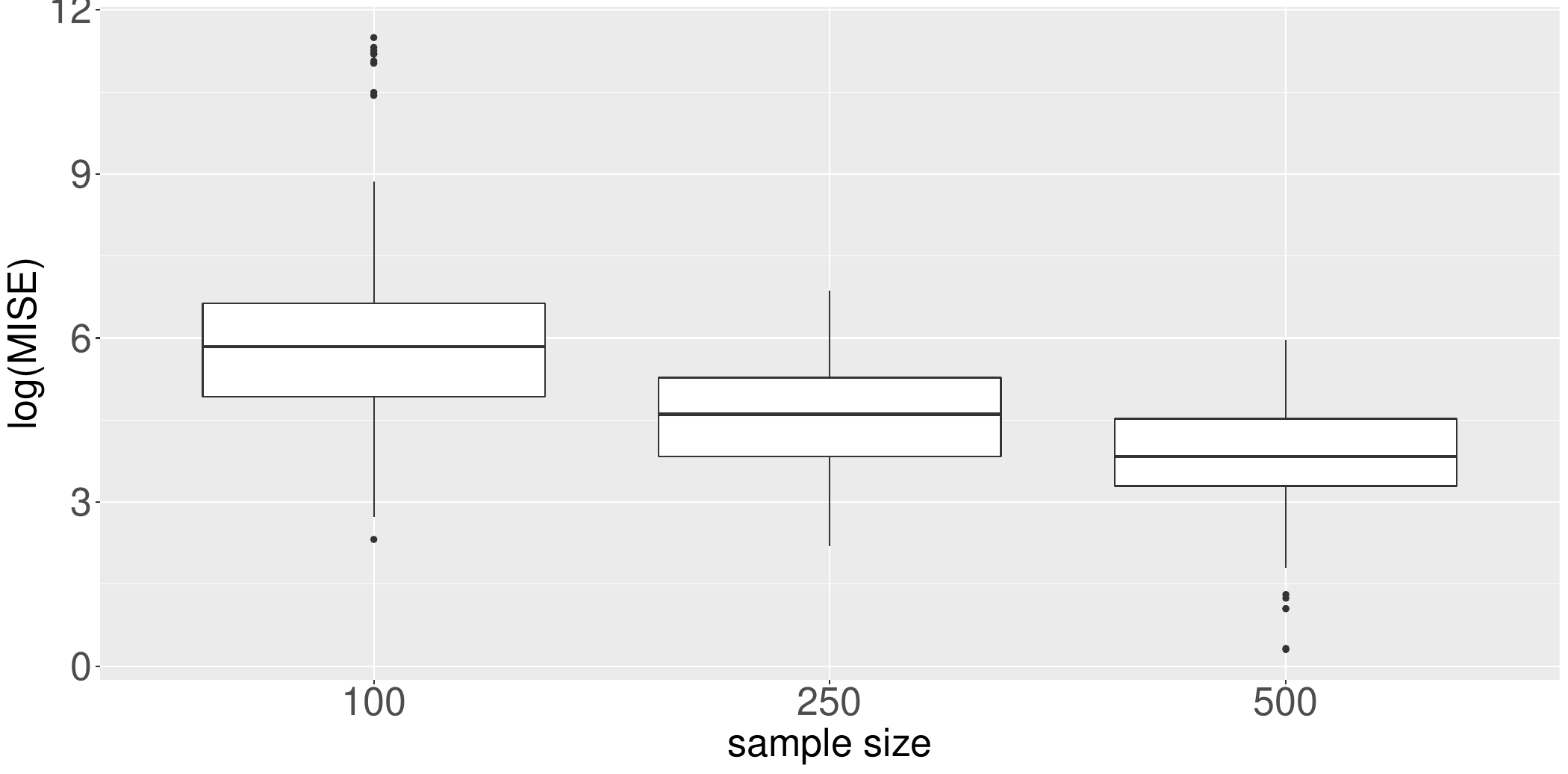} 
  \caption{\label{mise} \footnotesize Side-by-side boxplots of MISE {on the log scale} for Monte Carlo simulation study {for Scenario~1}.}
\end{figure}

\noindent We now move to the conditional density. To compare the fitted conditional density against the true as the sample size increases, we resort to the {mean integrated squared error} (MISE): 
\begin{equation*}
  \text{MISE} = \E\bigg[\int\int \{\widehat{f}(y \mid \mathbf{x}) - f(y \mid \mathbf{x})\}^2 \, \dif \mathbf{x} \, \dif y\bigg],
\end{equation*}
{where the expectation is taken so to summarize the average behavior of the randomness on the double integral that stems from the posterior mean estimate $\widehat{f}$ \citep[][Section~2.3]{wand1995}.}
Figure~\ref{mise} {shows} a boxplot {of} the MISE {of} each run of the simulation experiment {for Scenario~1}. {As can be} seen from Figure~\ref{mise}, MISE tends to decrease as the {sample size} increases, thus indicating that a better performance of the proposed methods is to be expected on larger samples; {the same holds for the remainder scenarios as can be seen from the boxplots of MISE available in the Supplementary Material}. {To examine the quality of each fit corresponding to a simulated dataset, $\{(\textbf{x}_i, y_i)\}_{i = 1}^n$, we use randomized quantile residuals \citep{dunn1996} adapted to our model, that is, $\{\varepsilon_i\}_{i = 1}^n = [\Phi^{-1}\{\widehat{F}(y_i|\mathbf{x}_i)\}]_{i = 1}^n$, where $\Phi^{-1}$ is the quantile function of the standard Normal distribution and $\widehat{F}$ is the posterior mean estimate of the conditional distribution function. Figure~\ref{residuals} depicts the 
  corresponding posterior mean QQ-plot of randomized quantile residuals against the theoretical standard
  normal quantiles, and it evidences an acceptably good fit of the model for Scenarios~1--4. } {Finally, to supplement the analysis, we also report in the Supplementary Material numerical experiments that tend to support the claim that the performance of the proposed methods is relatively satisfactory at separating covariate effects for the lower values and tail (Section 2.1 of Supplementary Material), and that this also holds in a misspecified setting. Despite the overall good performance, in all fairness not everything is perfect; specifically, our supporting numerical experiments reveal that: i) the correspondence between ``non-constant'' effects on lower quantiles and coefficients of $\kappa(\mathbf{x})$ is weaker for Scenario~3; ii) bias appears on the lower conditional quantile function estimates, under misspecification.} 


\begin{figure}[H]
	\centering
	\begin{minipage}{0.48\linewidth}\centering\hspace{.7cm}
		\textbf{Scenario~1}\\
		\footnotesize		(light effects for {lower values}, light effects for tail)\\
	\end{minipage}
	\begin{minipage}{0.48\linewidth}\centering\hspace{.7cm} 
		\textbf{Scenario~2}\\
		\footnotesize		(light effects for {lower values}, large effects for tail)\\
	\end{minipage} 
	\includegraphics[width=0.45\textwidth]{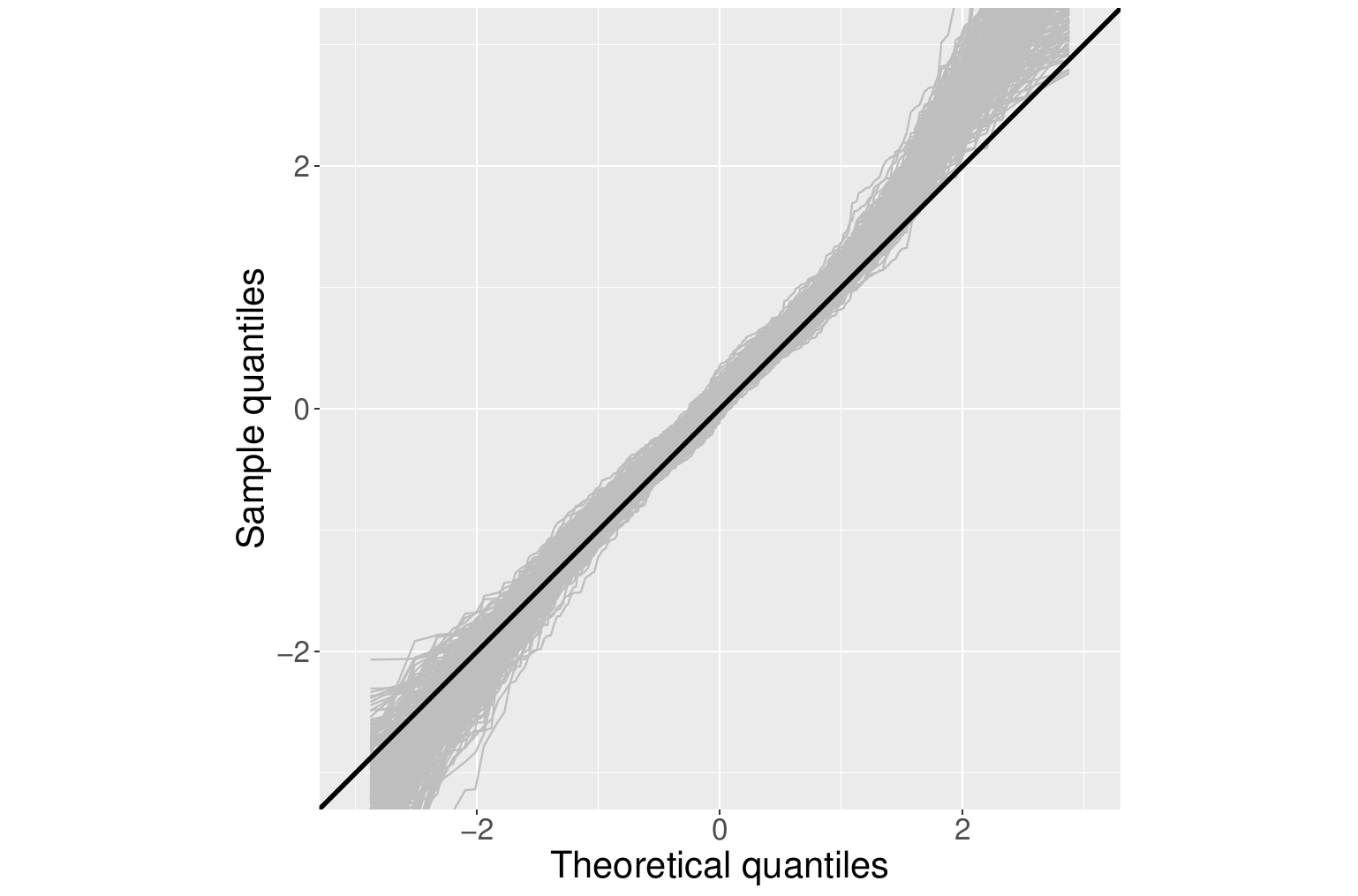} 
	\includegraphics[width=0.45\textwidth]{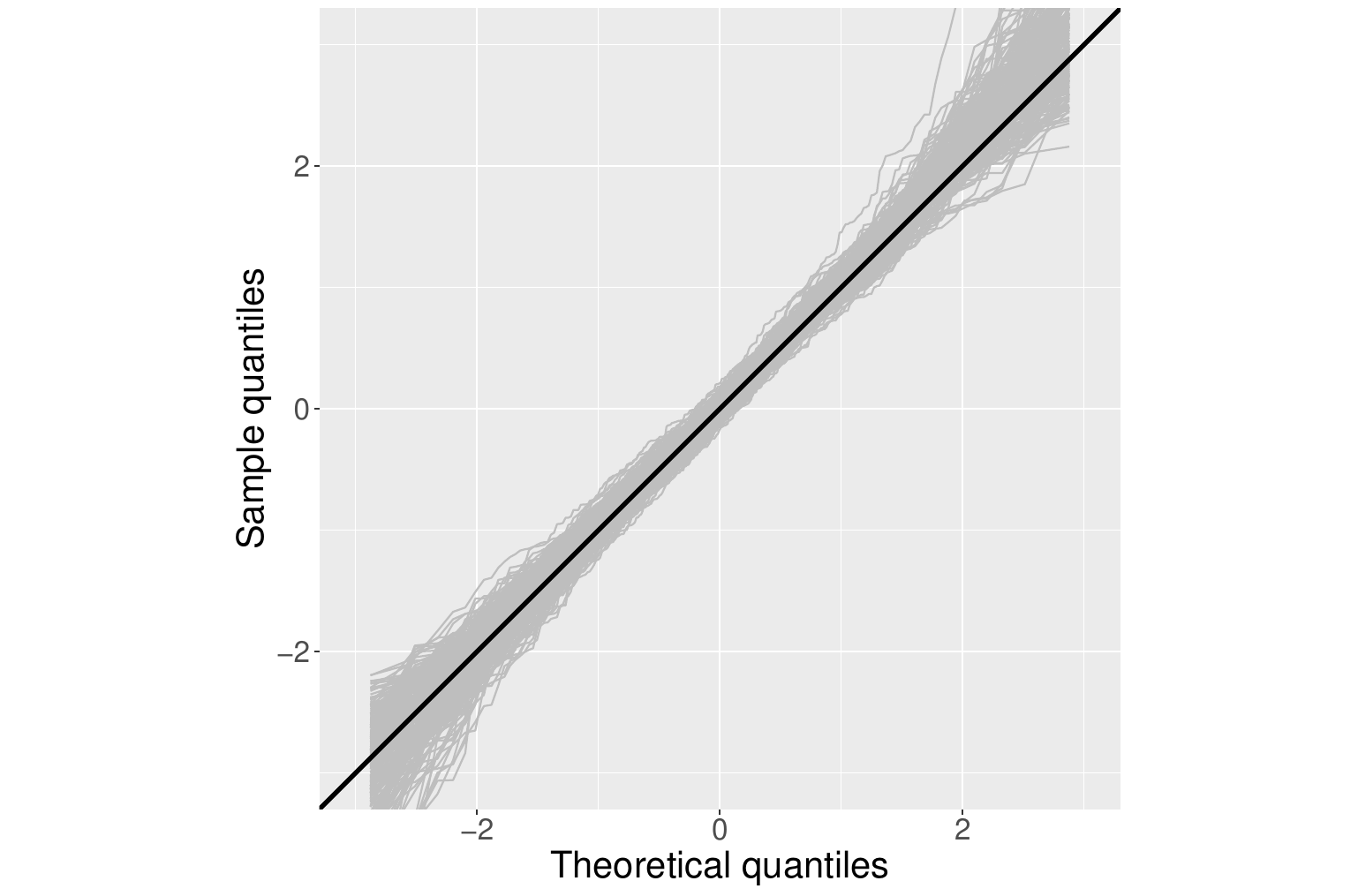}\\
	\vspace{0.5cm}
	\begin{minipage}{0.48\linewidth}\centering\hspace{.7cm}
		\textbf{Scenario~3}\\
		\footnotesize		(large effects for {lower values}, light effects for tail)\\
	\end{minipage}
	\begin{minipage}{0.48\linewidth}\centering\hspace{.7cm}
		\textbf{Scenario~4}\\
		\footnotesize		(large effects for {lower values}, large effects for tail)\\
	\end{minipage} 
	\includegraphics[width=0.45\textwidth]{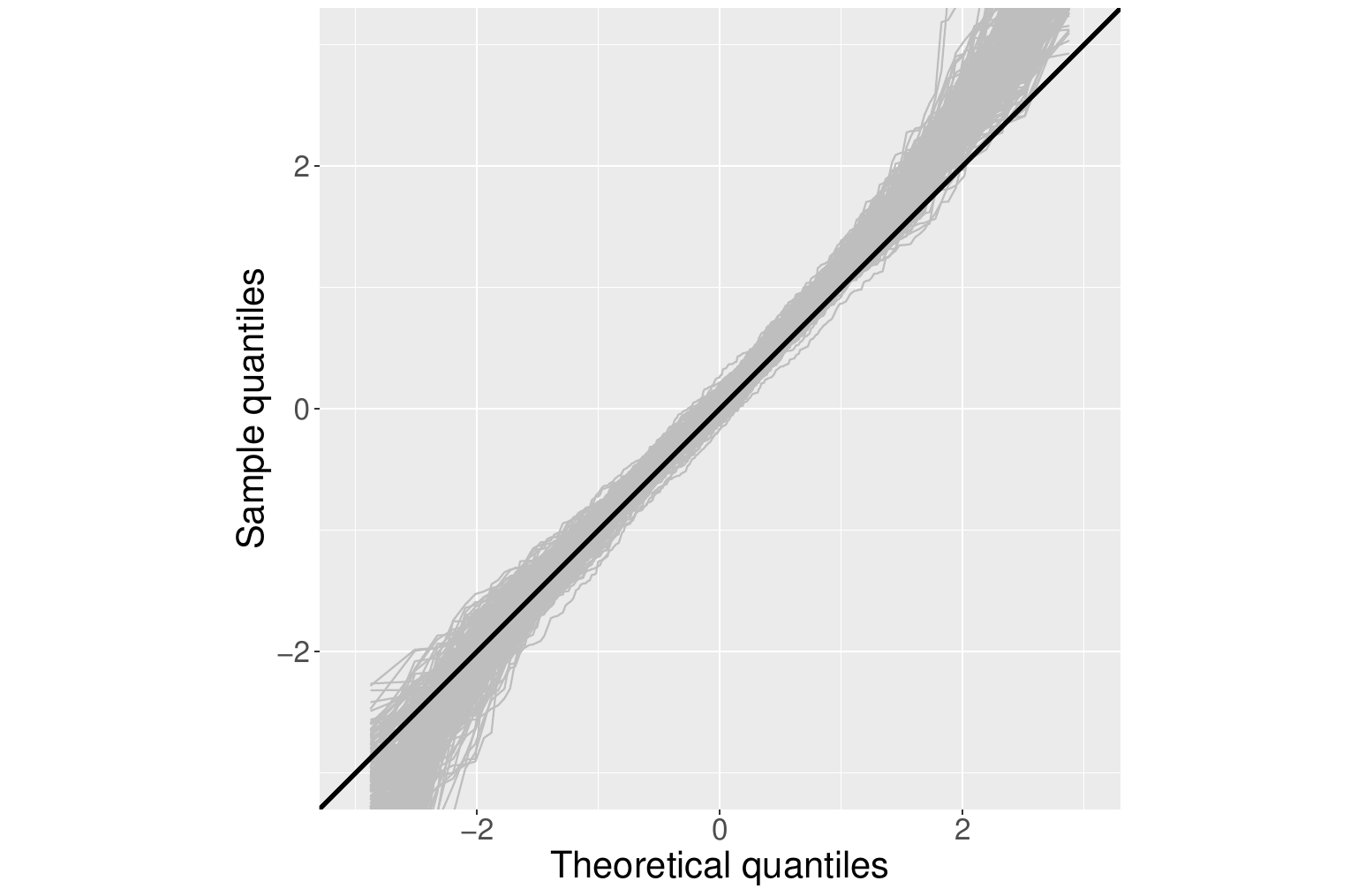}
	\includegraphics[width=0.45\textwidth]{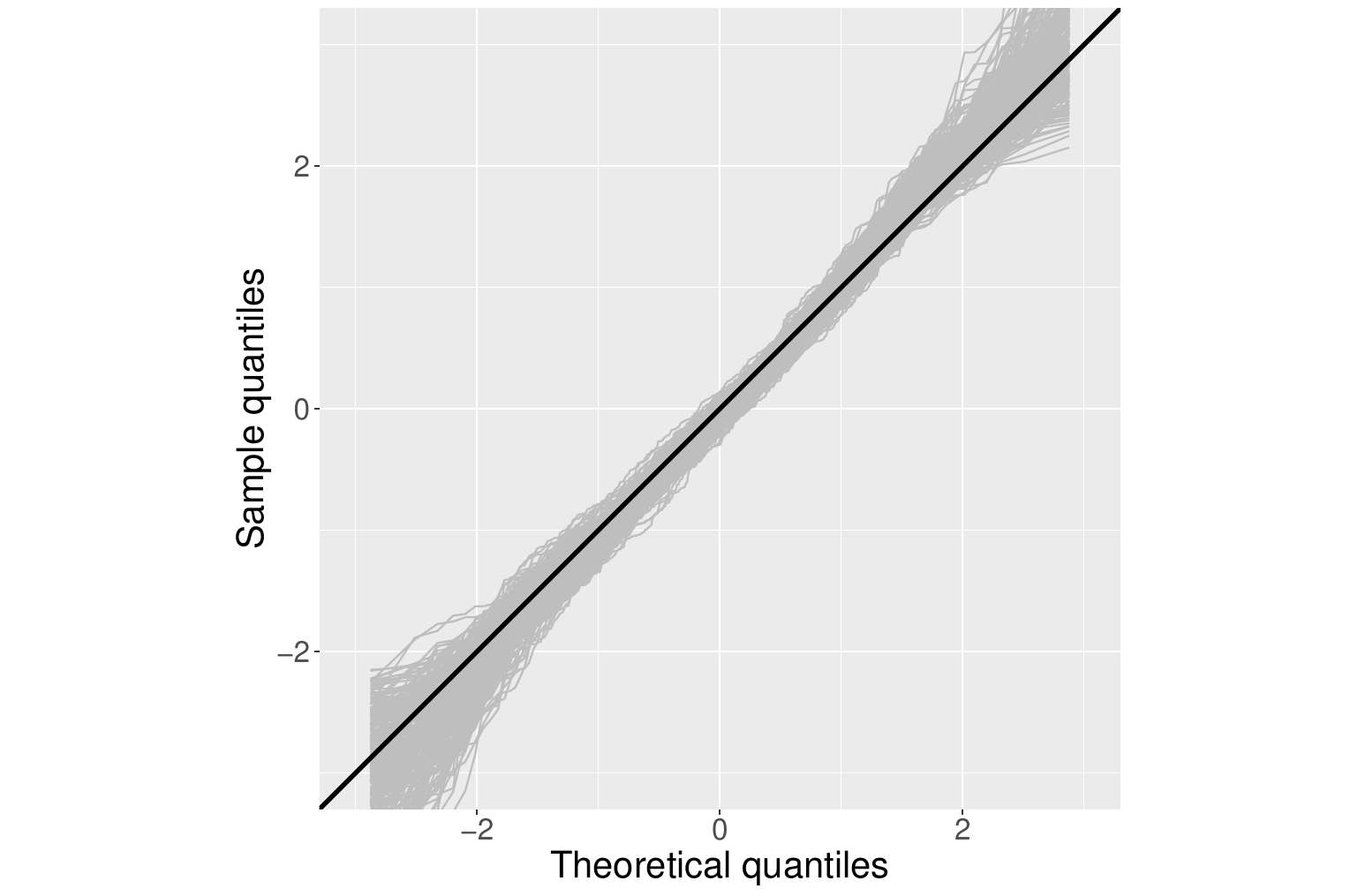}\\
	\caption{\label{rqrmc} \footnotesize {QQ-plots of randomized quantile residuals for Monte Carlo simulation study; each trajectory corresponds to a posterior mean QQ-plot from each simulated dataset.}}
\end{figure}

\section{\large\textsf{DRIVERS OF MODERATE AND EXTREME RAINFALL IN MADEIRA}}\label{application}
\subsection{\large\textsf{MOTIVATION, APPLIED CONTEXT, AND DATA DESCRIPTION}} 
{We now showcase the proposed methodology with a climatological
  illustration with data from Funchal, Madeira (Portugal); the island
  of Madeira is an archipelago of volcanic origin located in the
  Atlantic Ocean {about 900km} southwest of mainland Portugal. Prior to
  fitting the proposed model, we start by providing some background on
  the scientific problem of interest and by describing the data.
  Madeira has suffered a variety of extreme rainfall events over the
  last two centuries, including the flash floods of October 1803
  (800--1000 casualties) and those of February 2010---the latter with
  a death toll of 45 people \citep{fragoso2012, santos2017} and with
  an estimated damage of 1.4 billion {Euros} \citep{baioni2011}. Such
  violent rainfall events are often followed by damaging landslides
  events, including debris, earth, and mud flows.}

{To analyze such rainfall events in Funchal, Madeira, we have
  gathered data from the National Oceanic and Atmospheric Administration
  (\url{www.noaa.gov}). Specifically, we have gathered total monthly
  precipitation (.01 inches), as well as the following potential drivers
  for extreme rainfall: Atlantic multi-decadal Oscillation (AMO), El
  Ni\~no-Southern Oscillation (expressed by NINO34 index) (ENSO),
  North Pacific Index (NP), Pacific Decadal Oscillation (PDO),
  Southern Oscillation Index (SOI), and North Atlantic Oscillation
  (NAO). The sample period covers January 1973 to June 2018, thus
  including episodes such as the violent flash floods of 1979, 1993,
  2001, and 2010 \citep{baioni2011}. After eliminating the dry events
  (i.e.{,}~zero precipitation) and the missing precipitation data 
  (two observations), we are left with a total of 532 observations.}

{The potential drivers for extreme rainfall above have been
  widely examined in the climatological literature, mainly on large
  landmasses. In particular, it has been suggested that in North
  America ENSO, PDO, and NAO play a key role governing the occurrence
  of extreme rainfall events \citep{kenyon2010, zhang2010, whan2017};
  yet for the UK, while NAO is believed to impact the occurrence of
  extreme rainfall events, no influence of PDO nor AMO has been
  detected \citep{brown2018}. The many peculiarities surrounding
  Madeira climate (e.g.{,}~Azores {A}nticyclone, Canary {C}urrent, Gulf
  {S}tream, \textit{etc}{.}), along with the negative impact that flash
  floods and landslide events produce on the island, motivate us to
  ask: i) whether such drivers are relevant for explaining extreme
  rainfall episodes in Madeira; ii) whether such drivers are also
  relevant for moderate rainfall events.}

\subsection{\large\textsf{TRACKING DRIVERS OF MODERATE AND EXTREME RAINFALL}}
{One of the goals of the analysis is to use the lenses of our model so
to learn what are the drivers of moderate and extreme rainfall in
Funchal. To conduct such inquiry, we use {the
  full model from Section~\ref{extension} (see {Eq.~\eqref{links2}}), with power carrier  $G_{\kappa(\mathbf{x})}(v) = v^{\kappa(\mathbf{x})}$ and with link functions 
  \begin{equation*}
    \kappa(\mathbf{x}) = \exp(\mathbf{x}^{\T}\betab), \quad
    \nu(\mathbf{x}) = \exp(\mathbf{x}^{\T}\alphab), \quad
    \xi(\mathbf{x}) = \mathbf{x}^{\T} \gammab.
  \end{equation*}
  Covariates have been standardized {and} {we used} a flat Normal prior, $\text{N}(0, 100^2)$, for the intercepts,
and {uninformative Gamma priors, $\Ga(0.1, 0.1)$}, for the
{hyperparameters $\lambda_k$, $\lambda_\nu$, $\lambda_\xi$; here an identity link was used for 
  $\xi(\mathbf{x})$ as fitting a GPD to the response via likelihood methods suggested no compelling evidence
  in favor of an heavy-tailed response.}
After a burn-in period of {10\,000}
iterations, we collected a total of 20\,000 posterior samples. The
results of MCMC convergence diagnostics were satisfactory; in
particular, the effective sample sizes {were} acceptably high, ranging
from about {1000 to 5\,000}, and all values of Geweke's diagnostic {were}
satisfactory, ranging from about {$-3.2$ to $3.0$}. Figure~\ref{bands}
depicts the {credible intervals} for each regression coefficient.}
{To assess the fit of the proposed model we resort once more to randomized
  quantile residuals \citep{dunn1996}. Figure~\ref{residuals} evidences an acceptably good fit of the model;} {while the pointwise bands in Figure \ref{residuals} are narrow, they result from acceptably high effective sample sizes as mentioned above.}\begin{figure}
  \begin{minipage}{0.32\linewidth}
    \centering
    \includegraphics[scale = .22]{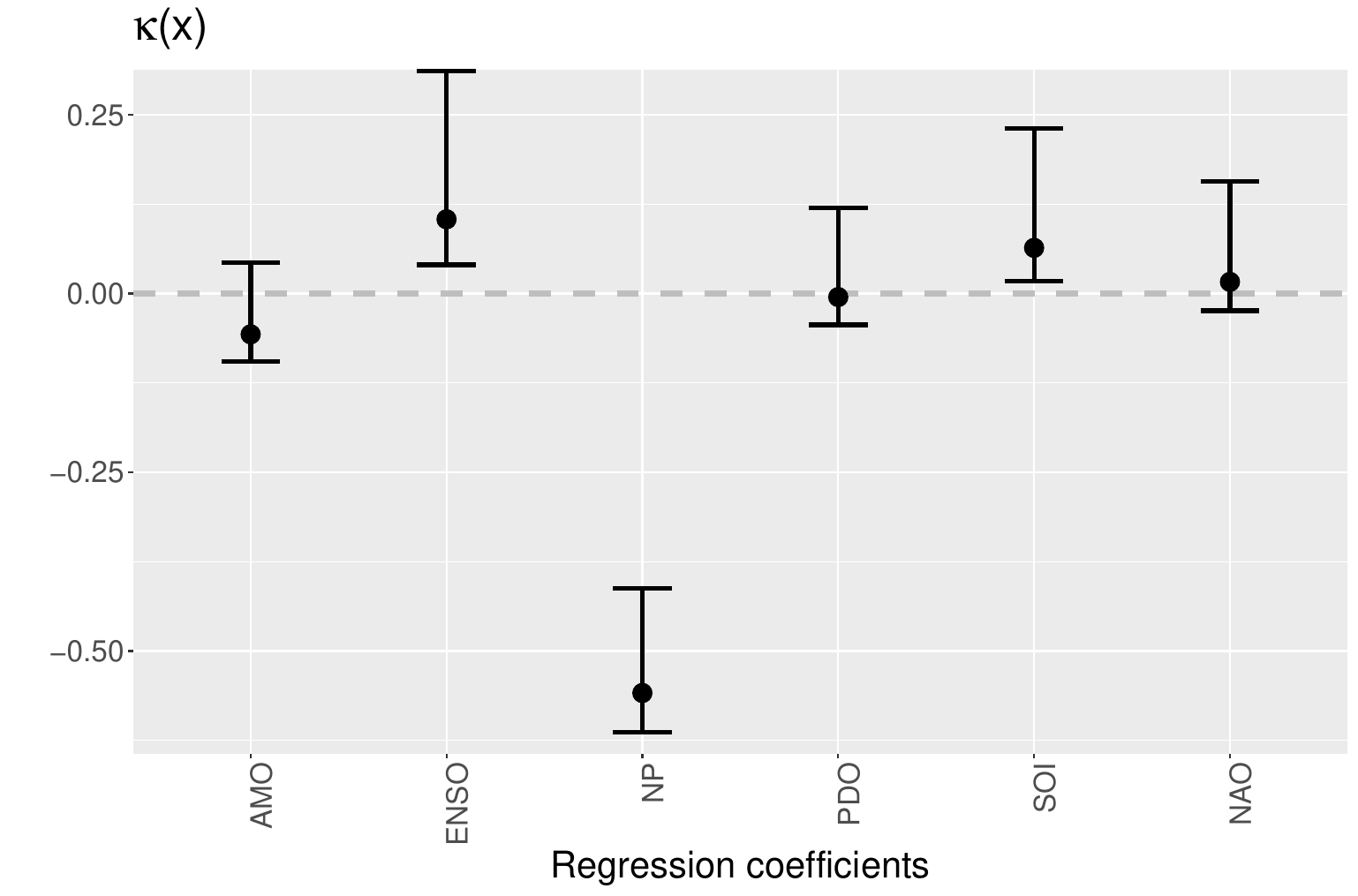}
  \end{minipage}
    \begin{minipage}{0.32\linewidth}
    \centering
    \includegraphics[scale = .22]{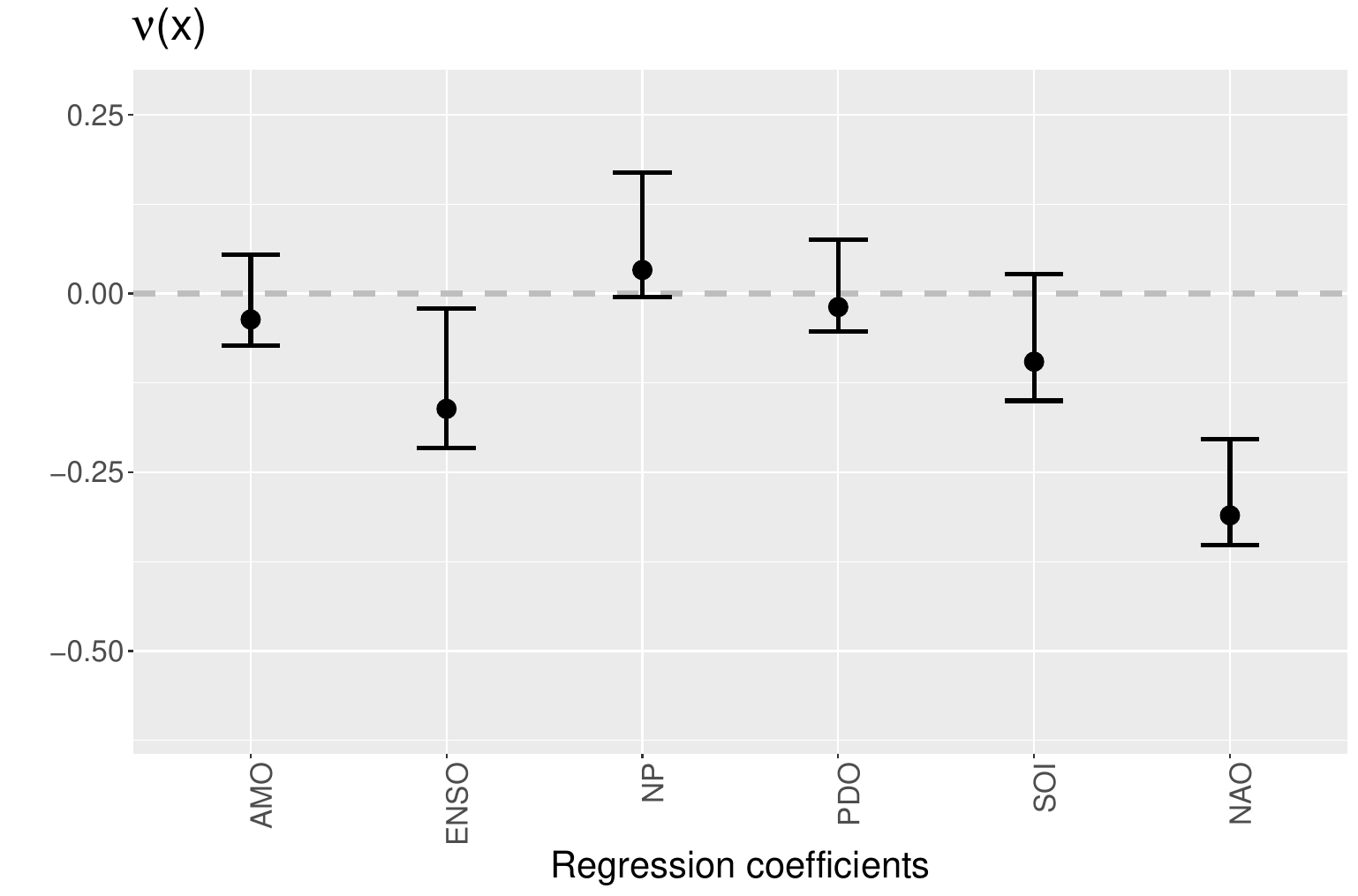}
  \end{minipage}
  \begin{minipage}{0.32\linewidth}
    \centering
    \includegraphics[scale = .22]{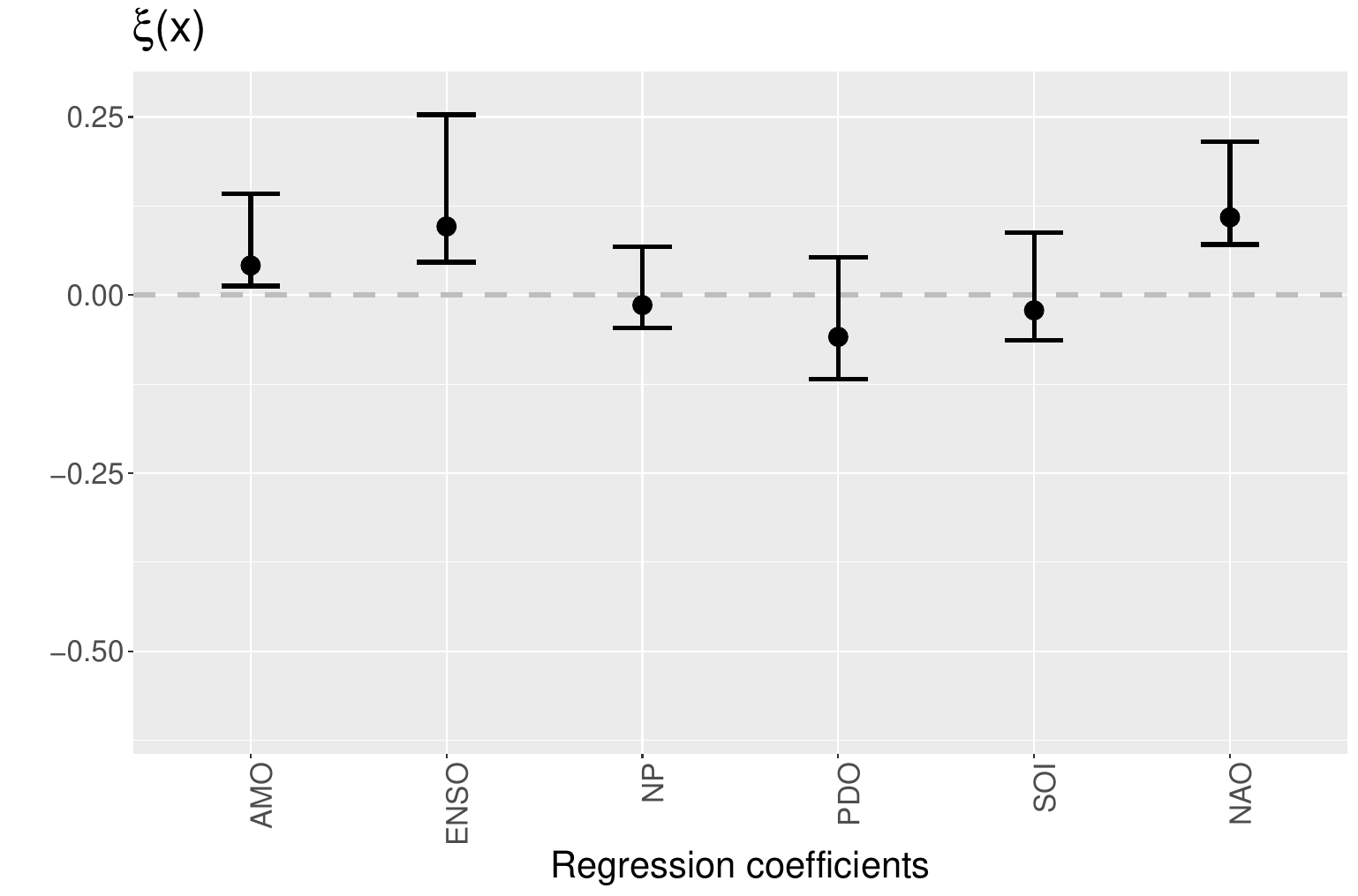}
  \end{minipage}
  \caption{\label{bands} \footnotesize {Credible intervals} for regression coefficients
    for inverse-link functions; the dots ($\bullet$) represent the
    posterior means and the dashed line represents the reference line
    $\beta = 0$.}
\end{figure}

\begin{figure}
  \centering
  \includegraphics[scale = .3]{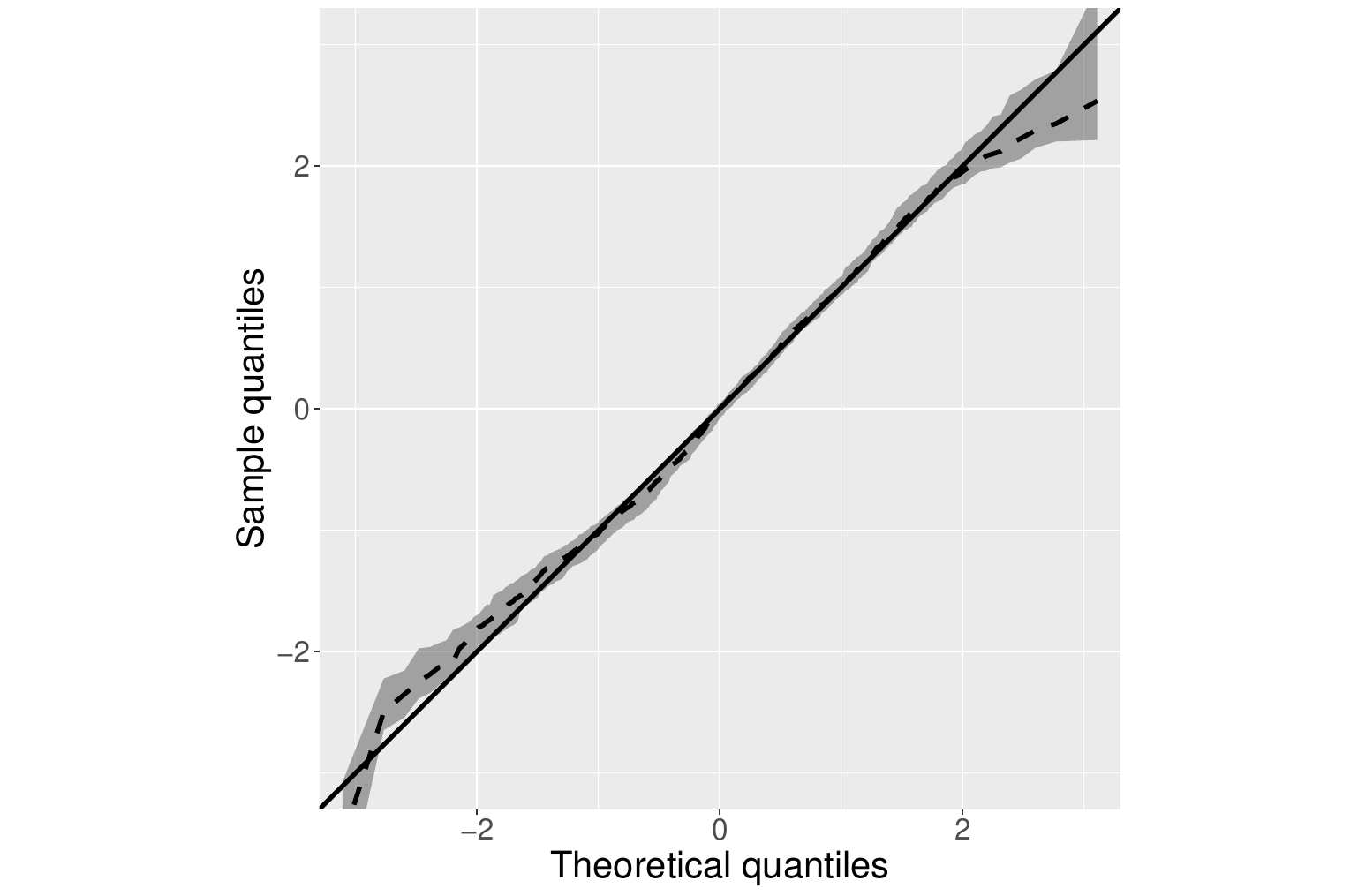} 
  \caption{\label{residuals} \footnotesize {QQ-plot of} randomized quantile residuals; the dashed line represents the posterior mean plotted along with {pointwise} credible bands.}
\end{figure}
{{Figure~\ref{bands}} {(middle and right panels)} {suggests} that the key drivers for extreme
  rainfall in Funchal are NAO and ENSO, along with a possible modest
  influence of AMO on the magnitude of extreme rainfall events. To put
  it differently, such results hint that a higher NAO, ENSO, and AMO
  tend to be associated with extreme rainfall episodes in
  Funchal. Such findings are thus reasonably in line with those of
  \citet{kenyon2010} \citet{zhang2010}, and \citet{whan2017} for North
  America. Yet, similarly to \citet{brown2018}, we find no relevant
  impact of PDO, {given the rest}, on extreme rainfall spells.} {Interestingly, of
  all these potencial drivers for extreme rainfall, {only ENSO}
  seems to be statistically associated with moderate rainfall; see
  Figure~\ref{bands} (left). Also, NP is the most relevant driver of
  moderate rainfall, and yet the analysis suggests it plays no role on
  extreme rainfall. Additionally, Figure~\ref{bands} (left) suggests
  that an increase in NP leads {to a heavier} left tail (i.e.{,}~dryer
  months), whereas an increase in ENSO leads to less mass close to
  zero (i.e.{,}~rainy months). {Finally, to supplement the analysis,  
  we also present conditional quantiles in the Supplementary Material
  so to directly assess how rainfall itself is impacted by all these
  potential drivers.
}

\section{\large\textsf{CLOSING REMARKS}}\label{discussion}
In this paper we have introduced a Lasso-type model for the {lower values} and
tail of a possibly heavy-tailed response. The proposed model: i) bypasses
the need for threshold selection ($u_{\mathbf{x}}$); ii) models both the
conditional {lower values} and conditional tails; iii) it is naturally tailored
for variable selection. {In addition, contrarily to GPD-based approach of \cite{davison1990} the proposed model does not suffer from lack of threshold stability, as indeed our model does not depend on a threshold; the fact that the Davison--Smith approach does not retain the threshold stability property is known since \citep[][Section~2.2]{eastoe2009}, and it implies that selection of covariates in the parameters of the GPD is not invariant to threshold choice.} {Yet, other approaches that circumvent the threshold stability issue are available beyond our model, including an inhomogeneous point process formulation for extremes \citep[][Chapter 7]{coles2001}.} Interestingly, the proposed model can be regarded as a Lasso-type model for quantile 
regression that is tailored for both the {lower values} and for extrapolating
into the tails. As a byproduct, our paper has {links} with the Bayesian
literature on Bayesian distributional regression \citep{umlauf2018},
and with the recent paper {of} \cite{groll2019}.

{Despite the considerations above on threshold selection there
  is no `free lunch'; indeed, here the parameters of the model and the
  choice of the carrier dictate the behavior of the quantiles in the
  whole range of the distribution, while the impact of some parameters
  is smaller in the lower or upper tail. Yet, competing
  EGPD models that stem from different choices for the carrier function can be
  formally compared and ranked via standard model selection criteria, such
  as Log Pseudo Marginal Likelihood \citep[e.g.][]{geisser1979, gelfand1994}.}

Some final comments on future research are in order.  A version of our
model that includes additionally a regression model for point masses
at zero would be natural for a variety of contexts, such as for
modeling long periods without rain or droughts. Semicontinuous
responses that consist of a mixture of {zeros} and a continuous positive
distribution are indeed common in practice
\citep[e.g.{,}][]{olsen2001}. 
Another natural avenue for future research would endow the model
with further flexibility by resorting to a 
generalized additive model, where the smooth function of each
covariate is modeled using B-spline basis functions.  The latter
extension would however require a group lasso \citep{yuan2006}, 
shrinking groups of regression coefficients (per smooth function) towards
zero. While {we focus here} on modeling positive random variables, another
interesting extension of the proposed model would consider the left
{endpoint} itself as a parameter {rather} than fixing it at zero. 
Finally, from an inference perspective it would also seem natural
defining a semiparametric Bayesian version of our model that would set
a prior directly on the covariate-specific carrier function $G_{\mathbf{x}}$ {rather than specifying a power carrier function in advance, }and to model the conditional density as $F(y \mid \mathbf{x})=G_{\mathbf{x}}{\{H(y \mid \mathbf{x})\}}$. This approach would however require setting a prior on the space $\mathscr{G}$ of all carrier functions so {as} to ensure that $G_{\mathbf{x}}$ obeys Assumptions~{A--C}, for all $\mathbf{x}$.  While priors on
spaces of functions are an active field of research
\citep{ghosal2015}, definition of priors over $\mathscr{G}$ is to our
knowledge an open problem; {a natural line of attack to
construct such a prior would be via   
random Bernstein polynomials \citep{petrone2002}, with weight
constraints derived from Lemma~1 of \cite{tencaliec2019}; {indeed, by modeling the carrier
  function with a Bernstein polynomial basis a higher level of flexibility could be
  achieved in comparison to that offered by Examples~\ref{powerc}--\ref{powerm}.
}}

\section*{\large\textsf{ACKNOWLEDGMENTS}} \label{acknowledgements}\footnotesize
We are grateful to the Editor, the Associate Editor, and two Referees for their insightful and constructive re- marks on an earlier version of the paper. We thank participants of IWSM (International Workshop on Statistical Modeling) 2019 and of EVA 2019 for constructive comments and discussions. We also thank, without implicating, Vanda In\'acio de Carvalho, Ioannis Papastathopoulos, Philippe Naveau, Ant\'onia Turkman, and Feridun Turkman for helpful comments and fruitful discussions. This work was partially supported by FCT (Funda\c c\~ao para a Ci\^encia e a Tecnologia, Portugal), through the projects PTDC/MAT-STA/28649/2017 and UID/MAT/00006/2020.

\renewcommand\refname{\textsf{REFERENCES}} 
\bibliographystyle{asa2.bst} 
\bibliography{library.bib}

\begin{thebibliography}{60}
\newcommand{\enquote}[1]{``#1''}
\expandafter\ifx\csname natexlab\endcsname\relax\def\natexlab#1{#1}\fi

\bibitem[{Arnold and Press(1989)}]{arnold1989}
Arnold, B.~C. and Press, S.~J. (1989), Bayesian estimation and prediction for
  Pareto data, \textit{Journal of the American Statistical Association}, 84,
  1079--1084.

\bibitem[{Baioni et~al.(2011)Baioni, Castaldini, and Cencetti}]{baioni2011}
Baioni, D., Castaldini, D., and Cencetti, C. (2011), Human activity and
  damaging landslides and floods on Madeira Island. \textit{Natural Hazards \&
  Earth System Sciences}, 11, 3035--3046.

\bibitem[{Behrens et~al.(2004)Behrens, Lopes, and Gamerman}]{behrens2004}
Behrens, C.~N., Lopes, H.~F., and Gamerman, D. (2004), Bayesian analysis of
  extreme events with threshold estimation, \textit{Statistical Modelling}, 4,
  227--244.

\bibitem[{Beirlant et~al.(2004)Beirlant, Goegebeur, Segers, and
  Teugels}]{beirlant2004}
Beirlant, J., Goegebeur, Y., Segers, J., and Teugels, J. (2004),
  \textit{Statistics of {{Extremes}}: {{Theory}} and {{Applications}}}, Wiley,
  Hoboken, NJ: {Wiley}.

\bibitem[{Bhattacharya et~al.(2015)Bhattacharya, Pati, Pillai, and
  Dunson}]{bhat2015}
Bhattacharya, A., Pati, D., Pillai, N.~S., and Dunson, D.~B. (2015),
  Dirichlet--Laplace priors for optimal shrinkage, \textit{Journal of the
  American Statistical Association}, 110, 1479--1490.

\bibitem[{Brown(2018)}]{brown2018}
Brown, S.~J. (2018), The drivers of variability in UK extreme rainfall,
  \textit{International Journal of Climatology}, 38, e119--e130.

\bibitem[{Cabras and Castellanos(2011)}]{cabras2011}
Cabras, S. and Castellanos, M.~E. (2011), A Bayesian approach for estimating
  extreme quantiles under a semiparametric mixture model, \textit{ASTIN
  Bulletin}, 41, 87--106.

\bibitem[{Carreau and Bengio(2009)}]{carreau2009}
Carreau, J. and Bengio, Y. (2009), A hybrid {{Pareto}} model for asymmetric
  fat-tailed data: The univariate case, \textit{Extremes}, 12, 53--76.

\bibitem[{Carvalho et~al.(2010)Carvalho, Polson, and Scott}]{carvalho2010}
Carvalho, C.~M., Polson, N.~G., and Scott, J.~G. (2010), The horseshoe
  estimator for sparse signals, \textit{Biometrika}, 97, 465--480.

\bibitem[{Castellanos and Cabras(2007)}]{castellanos2007}
Castellanos, M.~E. and Cabras, S. (2007), A default Bayesian procedure for the
  generalized Pareto distribution, \textit{Journal of Statistical Planning and
  Inference}, 137, 473--483.

\bibitem[{Chavez-Demoulin and Davison(2005)}]{chavez-demoulin2005}
Chavez-Demoulin, V. and Davison, A.~C. (2005), Generalized additive modelling
  of sample extremes, \textit{Journal of the Royal Statistical Society: Series
  C (Applied Statistics)}, 54, 207--222.

\bibitem[{Chavez-Demoulin et~al.(2016)Chavez-Demoulin, Embrechts, and
  Hofert}]{chavez-demoulin2016}
Chavez-Demoulin, V., Embrechts, P., and Hofert, M. (2016), An extreme value
  approach for modeling operational risk losses depending on covariates,
  \textit{Journal of Risk and Insurance}, 83, 735--776.

\bibitem[{Chernozhukov(2005)}]{chernozhukov2005}
Chernozhukov, V. (2005), Extremal quantile regression, \textit{Annals of
  Statistics}, 33, 806--839.

\bibitem[{Coles(2001)}]{coles2001}
Coles, S. (2001), \textit{An {{Introduction}} to {{Statistical Modeling}} of
  {{Extreme Values}}}, London: {Springer}.

\bibitem[{Davison and Smith(1990)}]{davison1990}
Davison, A.~C. and Smith, R.~L. (1990), Models for exceedances over high
  thresholds {(with Discussion)}, \textit{Journal of the Royal Statistical
  Society: Series B (Statistical Methodology)}, 393--442.

\bibitem[{de~Zea~Bermudez and Kotz(2010)}]{de2010}
de~Zea~Bermudez, P. and Kotz, S. (2010), Parameter estimation of the
  generalized Pareto distribution—Part II, \textit{Journal of Statistical
  Planning and Inference}, 140, 1374--1388.

\bibitem[{de~Zea~Bermudez and Turkman(2003)}]{de2003}
de~Zea~Bermudez, P. and Turkman, M.~A. (2003), Bayesian approach to parameter
  estimation of the generalized Pareto distribution, \textit{Test}, 12,
  259--277.

\bibitem[{do~Nascimento et~al.(2012)do~Nascimento, Gamerman, and
  Lopes}]{nascimento2012}
do~Nascimento, F.~F., Gamerman, D., and Lopes, H.~F. (2012), A semiparametric
  Bayesian approach to extreme value estimation, \textit{Statistics and
  Computing}, 22, 661--675.

\bibitem[{Dunn and Smyth(1996)}]{dunn1996}
Dunn, P.~K. and Smyth, G.~K. (1996), Randomized quantile residuals,
  \textit{Journal of Computational and Graphical Statistics}, 5, 236--244.

\bibitem[{Eastoe and Tawn(2009)}]{eastoe2009}
Eastoe, E.~F. and Tawn, J.~A. (2009), Modelling non-stationary extremes with
  application to surface level ozone, \textit{Journal of the Royal Statistical
  Society: Series C (Applied Statistics)}, 58, 25--45.

\bibitem[{Embrechts et~al.(1997)Embrechts, Kl{\"u}ppelberg, and
  Mikosch}]{embrechts1997}
Embrechts, P., Kl{\"u}ppelberg, C., and Mikosch, T. (1997), \textit{Modelling
  {{Extremal Events}} for {{Insurance}} and {{Finance}}}, New York: {Springer}.

\bibitem[{Fragoso et~al.(2012)Fragoso, Trigo, Pinto, Lopes, Lopes, Ulbrich, and
  Magro}]{fragoso2012}
Fragoso, M., Trigo, R., Pinto, J., Lopes, S., Lopes, A., Ulbrich, S., and
  Magro, C. (2012), The 20 February 2010 Madeira flash-floods: synoptic
  analysis and extreme rainfall assessment, \textit{Natural Hazards and Earth
  System Sciences}, 12, 715--730.

\bibitem[{Frigessi et~al.(2002)Frigessi, Haug, and Rue}]{frigessi2002}
Frigessi, A., Haug, O., and Rue, H. (2002), A dynamic mixture model for
  unsupervised tail estimation without threshold selection, \textit{Extremes},
  5, 219--235.

\bibitem[{Geisser and Eddy(1979)}]{geisser1979}
Geisser, S. and Eddy, W.~F. (1979), A predictive approach to model selection,
  \textit{Journal of the American Statistical Association}, 74, 153--160.

\bibitem[{Gelfand and Dey(1994)}]{gelfand1994}
Gelfand, A.~E. and Dey, D.~K. (1994), Bayesian model choice: asymptotics and
  exact calculations, \textit{Journal of the Royal Statistical Society: Series
  B (Statistical Methodology)}, 56, 501--514.

\bibitem[{Gelman et~al.(2015)Gelman, Lee, and Guo}]{gelman2015}
Gelman, A., Lee, D., and Guo, J. (2015), \texttt{Stan}: A probabilistic
  programming language for Bayesian inference and optimization,
  \textit{Journal of Educational and Behavioral Statistics}, 40, 530--543.

\bibitem[{George and McCulloch(1993)}]{george1993}
George, E.~I. and McCulloch, R.~E. (1993), Variable selection via Gibbs
  sampling, \textit{Journal of the American Statistical Association}, 88,
  881--889.

\bibitem[{Ghosal and {Van der Vaart}(2015)}]{ghosal2015}
Ghosal, S. and {Van der Vaart}, A.~W. (2015), \textit{Fundamentals of
  Nonparametric {{Bayesian}} Inference}, {Cambridge University Press,
  Cambridge}.

\bibitem[{Groll et~al.(2019)Groll, Hambuckers, Kneib, and Umlauf}]{groll2019}
Groll, A., Hambuckers, J., Kneib, T., and Umlauf, N. (2019), LASSO-type
  penalization in the framework of generalized additive models for location,
  scale and shape, \textit{Computational Statistics \& Data Analysis}, 140,
  59--73.

\bibitem[{Huser and Genton(2016)}]{huser2016}
Huser, R. and Genton, M.~G. (2016), Non-stationary dependence structures for
  spatial extremes, \textit{Journal of Agricultural, Biological, and
  Environmental Statistics}, 21, 470--491.

\bibitem[{Kenyon and Hegerl(2010)}]{kenyon2010}
Kenyon, J. and Hegerl, G.~C. (2010), Influence of modes of climate variability
  on global precipitation extremes, \textit{Journal of Climate}, 23,
  6248--6262.

\bibitem[{Kleiber and Kotz(2003)}]{kleiber2003}
Kleiber, C. and Kotz, S. (2003), \textit{Statistical Size Distributions in
  Economics and Actuarial Sciences}, New York: Wiley.

\bibitem[{Koenker(2005)}]{koenker2005}
Koenker, R. (2005), \textit{Quantile {{Regression}}}, Cambridge, MA: {Cambridge
  University Press}.

\bibitem[{Koenker and Bassett(1978)}]{koenker1978}
Koenker, R. and Bassett, G. (1978), Regression quantiles,
  \textit{Econometrica}, 33--50.

\bibitem[{MacDonald et~al.(2011)MacDonald, Scarrott, Lee, Darlow, Reale, and
  Russell}]{macdonald2011}
MacDonald, A., Scarrott, C., Lee, D., Darlow, B., Reale, M., and Russell, G.
  (2011), A flexible extreme value mixture model, \textit{Computational
  Statistics \& Data Analysis}, 55, 2137--2157.

\bibitem[{Mahmoud and Abd El-Ghafour(2015)}]{mahmoud2015}
Mahmoud, M.~R. and Abd El-Ghafour, A.~S. (2015), Fisher information matrix for
  the generalized Feller--Pareto distribution, \textit{Communications in
  Statistics---Theory and Methods}, 44, 4396--4407.

\bibitem[{Naveau et~al.(2016)Naveau, Huser, Ribereau, and Hannart}]{naveau2016}
Naveau, P., Huser, R., Ribereau, P., and Hannart, A. (2016), Modeling jointly
  low, moderate, and heavy rainfall intensities without a threshold selection,
  \textit{Water Resources Research}, 52, 2753--2769.

\bibitem[{Olsen and Schafer(2001)}]{olsen2001}
Olsen, M.~K. and Schafer, J.~L. (2001), A two-part random-effects model for
  semicontinuous longitudinal data, \textit{Journal of the American
  Statistical Association}, 96, 730--745.

\bibitem[{Papastathopoulos and Tawn(2013)}]{papas2013}
Papastathopoulos, I. and Tawn, J.~A. (2013), Extended generalised Pareto models
  for tail estimation, \textit{Journal of Statistical Planning and Inference},
  143, 131--143.

\bibitem[{Park and Casella(2008)}]{park2008}
Park, T. and Casella, G. (2008), The Bayesian {Lasso}, \textit{Journal of the
  American Statistical Association}, 103, 681--686.

\bibitem[{Petrone and Wasserman(2002)}]{petrone2002}
Petrone, S. and Wasserman, L. (2002), Consistency of {{Bernstein}} polynomial
  posteriors, \textit{Journal of the Royal Statistical Society: Series B
  (Statistical Methodology)}, 64, 79--100.

\bibitem[{Plummer(2019)}]{plummer2019}
Plummer, M. (2019), {\texttt{rjags}}: Bayesian graphical models using MCMC. R
  package version 4-10, {\url{http://CRAN.R-project.org/package=rjags}}, .

\bibitem[{Polson and Sokolov(2019)}]{polson2019}
Polson, N.~G. and Sokolov, V. (2019), Bayesian regularization: From Tikhonov to
  horseshoe, \textit{Wiley Interdisciplinary Reviews: Computational
  Statistics}, e1463.

\bibitem[{Reich and Ghosh(2019)}]{reich2019}
Reich, B.~J. and Ghosh, S.~K. (2019), \textit{Bayesian Statistical Methods},
  Boca Raton, FL: Chapman \& Hall/CRC.

\bibitem[{Resnick(1971)}]{resnick1971}
Resnick, S.~I. (1971), Tail equivalence and its applications, \textit{Journal
  of Applied Probability}, 8, 136--156.

\bibitem[{Santos et~al.(2017)Santos, Santos, and Fragoso}]{santos2017}
Santos, M., Santos, J.~A., and Fragoso, M. (2017), Atmospheric driving
  mechanisms of flash floods in Portugal, \textit{International Journal of
  Climatology}, 37, 671--680.

\bibitem[{Schott(2016)}]{schott2016}
Schott, J.~R. (2016), \textit{Matrix Analysis for Statistics}, New York: Wiley.

\bibitem[{Stein(2020)}]{stein2020}
Stein, M.~L. (2020), Parametric models for distributions when interest is in
  extremes with an application to daily temperature, \textit{Extremes}, 1--31.

\bibitem[{Stein(2021)}]{stein2021}
--- (2021), A parametric model for distributions with flexible behavior in both
  tails, \textit{Environmetrics}, 32, e2658.

\bibitem[{Tencaliec et~al.(2019)Tencaliec, Favre, Naveau, Prieur, and
  Nicolet}]{tencaliec2019}
Tencaliec, P., Favre, A.-C., Naveau, P., Prieur, C., and Nicolet, G. (2019),
  Flexible semiparametric Generalized Pareto modeling of the entire range of
  rainfall amount, \textit{Environmetrics}, e2582.

\bibitem[{Tibshirani(1996)}]{tibshirani1996}
Tibshirani, R. (1996), Regression shrinkage and selection via the {Lasso},
  \textit{Journal of the Royal Statistical Society: Series B (Statistical
  Methodology)}, 58, 267--288.

\bibitem[{Umlauf and Kneib(2018)}]{umlauf2018}
Umlauf, N. and Kneib, T. (2018), A primer on Bayesian distributional
  regression, \textit{Statistical Modelling}, 18, 219--247.

\bibitem[{Villa(2017)}]{villa2017}
Villa, C. (2017), Bayesian estimation of the threshold of a generalised
  {Pareto} distribution for heavy-tailed observations, \textit{Test}, 26,
  95--118.

\bibitem[{Wand and Jones(1995)}]{wand1995}
Wand, M.~P. and Jones, M.~C. (1995), \textit{Kernel Smoothing}, London:
  {Chapman \& Hall}, 1st ed.

\bibitem[{Wang and Tsai(2009)}]{wang2009}
Wang, H. and Tsai, C.-L. (2009), Tail {{index regression}}, \textit{Journal of
  the American Statistical Association}, 104, 1233--1240.

\bibitem[{Whan and Zwiers(2017)}]{whan2017}
Whan, K. and Zwiers, F. (2017), The impact of ENSO and the NAO on extreme
  winter precipitation in North America in observations and regional climate
  models, \textit{Climate Dynamics}, 48, 1401--1411.

\bibitem[{Young and Smith(2005)}]{young2005}
Young, G.~A. and Smith, R.~L. (2005), \textit{Essentials of {{Statistical
  Inference}}}, Cambridge, UK: {Cambridge University Press}.

\bibitem[{Yuan and Lin(2006)}]{yuan2006}
Yuan, M. and Lin, Y. (2006), Model selection and estimation in regression with
  grouped variables, \textit{Journal of the Royal Statistical Society: Series
  B (Statistical Methodology)}, 68, 49--67.

\bibitem[{Zhang et~al.(2010)Zhang, Wang, Zwiers, and Groisman}]{zhang2010}
Zhang, X., Wang, J., Zwiers, F.~W., and Groisman, P.~Y. (2010), The influence
  of large-scale climate variability on winter maximum daily precipitation over
  North America, \textit{Journal of Climate}, 23, 2902--2915.

\bibitem[{Zhang et~al.(2020)Zhang, Naughton, Bondell, and Reich}]{zhang2020}
Zhang, Y.~D., Naughton, B.~P., Bondell, H.~D., and Reich, B.~J. (2020),
  Bayesian regression using a prior on the model fit: The R2-D2 shrinkage
  prior, \textit{Journal of the American Statistical Association}, 1--13.

\end{thebibliography}

\end{document}